\documentclass[aps,prl,twocolumn,groupedaddress,showpacs,preprintnumbers,
amsmath,amssymb]{revtex4}
\usepackage[dvips]{graphicx}
\usepackage{amsfonts}
\usepackage{amssymb}
\usepackage{float}
\usepackage{natbib}
\usepackage{amsmath}
\usepackage{float}
\usepackage{dcolumn}
\usepackage{lscape}
\usepackage{psfrag}   

\newcommand{\beq}{\begin{equation}}
\newcommand{\eeq}{\end{equation}}
\newcommand{\beqs}{\begin{equation*}}
\newcommand{\eeqs}{\end{equation*}}
\newcommand{\bfg}{\begin{figure}}
\newcommand{\efg}{\end{figure}}

\def\eq#1{(\ref{#1})}
\def\fig#1{Fig.~\ref{#1}}

\def\bea{\begin{eqnarray}}
\def\eea{\end{eqnarray}}

\def\ci{\mathrm{i}}
\def\ud{\mathrm{d}}

\def\rme{\mathrm{e}}

\def\v{\boldsymbol v}

\def\s{\boldsymbol s}
\def\u{\boldsymbol u}

\makeatletter
\newlength{\earraycolsep}
\def\eqnarray{\stepcounter{equation}\let\@currentlabel%
\theequation
\global\@eqnswtrue\m@th
\global\@eqcnt\z@\tabskip\@centering\let\\\@eqncr
$$\halign to\displaywidth\bgroup\@eqnsel\hskip\@centering
$\displaystyle\tabskip\z@{##}$&\global\@eqcnt\@ne
\hskip 2\earraycolsep \hfil$\displaystyle{##}$\hfil
&\global\@eqcnt\tw@ \hskip 2\earraycolsep
$\displaystyle\tabskip\z@{##}$\hfil
\tabskip\@centering&\llap{##}\tabskip\z@\cr}
\makeatother

\begin{document}

\title{Semiclassical Theory for Decay and Fragmentation Processes \\
                   in Chaotic Quantum Systems  
}

 \author{ Martha Guti\'errez}
\affiliation{Institut f\"ur Theoretische Physik, Universit\"at Regensburg, D-93040 Regensburg, Germany}
\author{Daniel Waltner}
\affiliation{Institut f\"ur Theoretische Physik, Universit\"at Regensburg, D-93040 Regensburg, Germany}
 \author{Jack Kuipers}
\affiliation{Institut f\"ur Theoretische Physik, Universit\"at Regensburg, D-93040 Regensburg, Germany}
 \author{Klaus Richter}
\affiliation{Institut f\"ur Theoretische Physik, Universit\"at Regensburg, D-93040 Regensburg, Germany}

\date{\today}

\begin{abstract} 
We consider quantum decay and photofragmentation processes in open chaotic systems in the 
semiclassical limit. We devise a semiclassical approach which allows us to consistently calculate quantum corrections to the classical decay to high order in an expansion in the inverse Heisenberg time. We present results for systems with and without time reversal symmetry and also for the symplectic case, as well as extending recent results to non-localized initial states.
We further analyze related photodissociation and photoionization phenomena and semiclassically compute cross-section correlations, including their Ehrenfest time dependence. 
\end{abstract}

\maketitle

\section{I. Introduction}
Physical phenomena involving decay processes have been addressed in many physical contexts. 
They play a central role in the study of excitation relaxation in semiconductor quantum dots 
and wires \cite{ref:Bacher99,ref:Kumar98}, photoionization via highly excited 
atomic \cite{ref:Stania05} or molecular \cite{ref:Baumert91} Rydberg states, 
photodissociation of molecules \cite{ref:Schinke93}, atoms in optically generated lattices and
cavities \cite{ref:atombilliard}, and optical micro-cavities \cite{ref:microcav}, 
to name a few examples.

For an open chaotic system it is well known that the classical probability of finding a particle 
inside the system at a certain time, the classical survival probability, decays exponentially in 
time, $\rho^{\rm cl}(t)= e^{-t/ \tau_d } $, where $\tau_d $ is the classical life or dwell time. 
Numerical calculations \cite{ref:Casati97}, however, revealed that the quantum survival 
probability deviates from the classical one at times comparable to $t^*\approx \sqrt{\tau_d t_H}$, 
where $t_H=2\pi\hbar/\Delta$ is the Heisenberg time, and $\Delta$ is the mean level spacing.
Theoretical calculations invoking supersymmetry techniques \cite{ref:Frahm97,ref:Savin97} 
confirmed these findings. There it could be shown that in the random matrix theory (RMT) limit,
the quantum decay $\rho(t)$ takes the form of a universal function, which only depends on the 
general symmetries of the system, the classical life time and the Heisenberg time. 
The first successful semiclassical approach to derive the RMT predictions for quantum graphs was 
performed in Ref.\ \cite{ref:Puhlmann05}, reproducing the first order RMT quantum corrections 
for networks with and without time-reversal symmetry. 

Recently, we have developed a semiclassical approach for calculating the 
decay of an initially localized wave function inside an arbitrary chaotic system  \cite{ref:Waltner08}.
The semiclassical framework used there involves correlated trajectories which have been shown to be
a powerful tool and the key to link classical hyperbolic dynamics with universal quantum
properties \cite{ref:Sieber01}. These semiclassical techniques have been recently extended and
widely applied in the context of level statistics
\cite{ref:chaos-general,ref:Heusler04,ref:Brouwer06B}, 
where multiple sums over periodic orbits have to be evaluated, 
as well as in the field of ballistic quantum transport involving Landauer-B\"uttiker formulas 
\cite{ref:Richter02,ref:Adagideli03,ref:Heusler06,ref:Brouwer06,ref:Ehrenfest2,ref:Jacquod06,ref:Kuipers08},
where trajectories start and end at the openings where the chaotic conductor is attached to leads. 
In Ref.\ \cite{ref:Waltner08} a unitarity problem  was encountered 
when using these semiclassical techniques to evaluate the contribution of pairs of interfering 
trajectories starting and ending inside the system. Therefore a new kind of diagram was considered, which is crucial for ensuring unitarity in problems involving open trajectories connecting two arbitrary points in the bulk. A similar type of trajectory appears in the semiclassical description of transport if the coupling between the chaotic conductor and the leads is not perfect, as shown in Ref. \cite{ref:Whitney07}.

In this article we generalize the approach presented in Ref.\ \cite{ref:Waltner08} for localized
initial wave functions to non-localized wave functions. We outline how to systematically obtain
higher order (in $t/t^\ast$) quantum corrections to the classical decay and present
terms up to the 7th order and 8th order, for systems with and without time reversal symmetry, respectively. 
We further calculate the survival probability for systems with spin-orbit interaction, corresponding 
to the symplectic RMT ensemble. 

Closely related to quantum decay are problems of atomic photoionization or molecular
photodissociation where the fragmentation mechanism involves photoexcitation to an 
intermediate excited resonant state (with corresponding complex classical dynamics)
which then subsequently decays by sending out a particle, i.e.\ an electron, atom or an ion.
In the semiclassical limit, spectral correlation functions for the related photoionization and  
photodissociation cross-sections can be expressed through the spectral form factor and the survival probability.
Earlier semiclassical treatments \cite{ref:Agam00,ref:Eckhardt00} of photo cross-sections were 
always limited through the diagonal approximation used which was relaxed in this context only
very recently \cite{ref:Waltner08}. Here we will present a detailed semiclassical treatment of 
the brief account given in \cite{ref:Waltner08} and extent the results by including Ehrenfest time 
effects for cross-section correlations and by computing higher order contributions.

This article is organized as follows: in Secs.\ II and III we present the semiclassical approach 
to the quantum survival probability, generalized to non-localized wave functions, by including a time average.
In Sec.\ IV this approach is further extended to derive higher order corrections for systems with 
and without time reversal symmetry as well as for the case of spin-orbit interaction which follows the 
universal RMT prediction for the symplectic case. In Sec.\ V we analyze fluctuations of the survival 
probability through its variance. In Sects.\ VI and VII we give a detailed semiclassical analysis of the
statistics of photofragmentation, including higher order corrections and the Ehrenfest time dependence 
of the leading quantum contributions. We conclude with an outlook in Sec.~VIII.


\section{II. Semiclassical approach to the survival probability}

The quantum mechanical survival probability as a measure of the decay is defined as 
\beq\label{qdec}
\rho\left(t\right)  = \int_A d{\bf r}\, \psi({\bf r},t)\psi^*({\bf r},t) \, ,
\eeq
where $\psi({\bf r},t)$ is a wave function and $A$ the volume of the system we are considering.
For a closed system $\rho(t)=1 $, while for an open system this no longer holds and $\rho(t)$ 
decays in time.  Expressing  $\psi({\bf r},t)$ in terms of the propagator $K({\bf r},{\bf r'},t)$,
\beq\label{tdeps}
\psi({\bf r},t)= \int_A d{\bf r'} K({\bf r},{\bf r'},t)\psi_0({\bf r'}) \, ,
\eeq
we have
\beq\label{qdec2}
\rho\left(t\right) = \int_A d{\bf r} d{\bf r'}  d{\bf r''} 
K({\bf r},{\bf r'},t)K^*({\bf r},{\bf r''},t) \psi_0({\bf r'})\psi_0^*({\bf r''}) \, ,
\eeq
where $\psi_0({\bf r})$ is the initial wave function at $t=0$. 

In order to calculate the semiclassical expression for $\rho\left(t\right)$, 
we replace the exact quantum propagator $K({\bf r},{\bf r'},t)$ with the semiclassical 
Van Vleck propagator \cite{ref:Gutzwiller90},
\beq\label{smcprop}
K^{\rm sc}\left(\mathbf{r},\mathbf{r'},t\right)=\frac{1}{(2\pi i\hbar)^{f/2}}\sum_{\tilde\gamma
\left(\mathbf{r'}\to\mathbf{r},t\right)}D_{\tilde\gamma }{\rm
e}^{\frac{i}{\hbar}S_{\tilde\gamma}(\mathbf{r},\mathbf{r'},t)}\, .
\eeq
Here $f$ is the dimension of the system (in the following we will consider $f=2$), 
$S_{\tilde \gamma}({\bf r},{\bf r'},t)=\int_{0}^{t}dt' L_{\tilde\gamma}
[\dot{\mathbf r}_{\tilde\gamma},{\mathbf r}_{\tilde\gamma},t']$  
is the classical Hamilton's principal function (with  $L_{\tilde \gamma}$ the Lagrangian) along the path $\tilde\gamma$ 
connecting ${\bf r}'$ and ${\bf r}$ in a time $t$, and $D_{\tilde\gamma} =\left|{\rm det}
\left(-\frac{\partial^2S_{\tilde \gamma}({\bf r},{\bf r'},t)}{\partial{\bf r}\partial{\bf r'}}\right)\right|^{1/2}
{\rm e}^{-i\frac{\pi}{2}\mu_{\tilde\gamma}}$ is the Van Vleck determinant including the Morse 
index $\mu_{\tilde\gamma}$.

The semiclassical survival probability is then given by
\bea\label{smcsp}
\rho^{\rm sc}(t)&=&\frac{1}{(2\pi \hbar)^{2}}
\int_A d{\mathbf r}d{\mathbf r}'d{\mathbf r}\psi_0({\mathbf r}')\psi_0^*({\mathbf r}'')\times \\  &&\times
\sum_{\tilde\gamma \left(\mathbf{r}'\to\mathbf{r},t\right)\atop \tilde\gamma' \left(\mathbf{r}''\to\mathbf{r},t\right) } D_{\tilde\gamma }D_{\tilde\gamma' }^*{\rm e}^{\frac{i}{\hbar}(S_{\tilde\gamma} -S_{\tilde\gamma'}) }.
\nonumber
\eea

In the following, we introduce a local time average in the survival probability which enables us
to neglect highly oscillating terms in the above double sum. We define
\beq
\bar{\rho}(t)\equiv\langle \rho^{\rm sc}(t) \rangle_{\Delta t}\equiv \frac{1}{\Delta t}\int_{t-\Delta
t/2}^{t+\Delta t/2}\rho^{\rm sc}(t')dt'
\eeq
with $\Delta t\ll t$. We will later see that for a localized initial wave packet 
$\bar{\rho}(t)\approx \rho(t)$ in the semiclassical limit, recalling the result of Ref.\ \cite{ref:Waltner08}. 

The phase difference in the double sum in Eq.\ \eq{smcsp} rapidly oscillates unless the two related
trajectories are correlated. Therefore most of the contributions will disappear due to the time average. 
The contributions that prevail the average are from pairs of correlated trajectories with action differences 
of the order of $\hbar$, which implies that the trajectories $\tilde\gamma$ and $\tilde\gamma'$ 
should be `similar'. This puts a restriction on the initial points of the two trajectories, i.e.\
they should be almost the same.  We can then expand trajectories $\tilde\gamma$ (or $\tilde\gamma'$) 
going from  $\mathbf{r}'$ (or $\mathbf{r}''$) to $\mathbf{r}$ in a time t around  trajectories $\gamma$ 
(or $\gamma'$) going from $\mathbf{r}_0=(\mathbf{r}'+\mathbf{r}'')/2$ to $\mathbf{r}$ in a time $t$. 
This expansion amounts to approximating the classical prefactors
$ D_{\tilde\gamma} \left( \mathbf{r},\mathbf{r}',t  \right)  
\approx D_{\gamma} \left( \mathbf{r},\mathbf{r}_0,t \right) $ and $ D_{\tilde\gamma'} 
\left( \mathbf{r},\mathbf{r}'',t  \right)  \approx D_{\gamma'} 
\left( \mathbf{r},\mathbf{r}_0,t \right)$, while expanding the phases in the exponents 
up to the first order, because the latter are more sensitive to small changes in their argument. 
The expansion of the actions yields
\beq
S_{\tilde\gamma} \left( \mathbf{r},\mathbf{r}',t \right) \approx S_{\gamma}
\left( \mathbf{r},\mathbf{r}_0,t  \right) - \frac{1}{2}\mathbf{q} \cdot \mathbf{p}_{\gamma,0}\, ,
\eeq
\beq
S_{\tilde\gamma'} \left( \mathbf{r},\mathbf{r}'',t \right) \approx S_{\gamma'}
\left( \mathbf{r},\mathbf{r}_0,t  \right) + \frac{1}{2}\mathbf{q} \cdot \mathbf{p}_{\gamma',0},
\eeq
where $\mathbf{q}=\mathbf{r}'-\mathbf{r}''$ and $\mathbf{p}_{\gamma,0}$ (or $\mathbf{p}_{\gamma',0}$) 
is the initial momentum of the trajectory $\gamma$ (or $\gamma'$). The semiclassical survival 
probability, Eq.~\eq{smcsp} then reads
\bea\label{eq4}
\bar{\rho}(t)&=& {\Big \langle}\frac{1}{(2\pi \hbar)^2}\int d{\mathbf r} d{\mathbf r}_0 d{\mathbf
q}\,\psi_0\left({\mathbf r}_0+\frac{{\mathbf q}}{2}\right)\psi_0^*\left({\mathbf r}_0-\frac{{\mathbf
q}}{2}\right)\times
\nonumber\\ 
& &\times\!\! \sum_{\gamma,\gamma'({\mathbf r}_0\to{\mathbf r},t)}D_{\gamma}D_{\gamma' }^*e^{\frac{i}{\hbar}(S_{\gamma} -S_{\gamma'})} {\rm e}^{-\frac{i}{\hbar} \mathbf{\bar p}^0_{\gamma\gamma'}\cdot\mathbf{q}}{\Big \rangle}_{\Delta t},
\eea
where $\mathbf{\bar p}^0_{\gamma\gamma'}=(\mathbf{p}_{\gamma,0}+\mathbf{p}_{\gamma',0})/2$.
This can be written as
\bea
\bar{\rho}(t)&=& {\Big \langle}\frac{1}{(2\pi \hbar)^2}\int d{\mathbf r} d{\mathbf r}_0 \times
\\\nonumber &&\times \sum_{\gamma,\gamma'({\mathbf r}_0\to{\mathbf r},t)}\!\!D_{\gamma}D_{\gamma' }^*e^{\frac{i}{\hbar}(S_{\gamma} -S_{\gamma'})} \rho_W\left({\mathbf r}_0,\mathbf{\bar p}^0_{\gamma\gamma'}\right){\Big \rangle}_{\Delta t},
\eea
 where
 \beq \rho_{W}({\mathbf r},{\mathbf p})=\int d{\mathbf r'}\psi_0\left({\mathbf r}+\frac{{\mathbf r'}}{2}\right)\psi_0^*\left({\mathbf r}-\frac{{\mathbf r'}}{2}\right){\rm e}^{-\frac{i}{\hbar}\mathbf{r}'\cdot \mathbf{p}},
\eeq
is the Wigner transformation of $\psi_0({\mathbf r}) $.  For an initial coherent state, the integrals 
over ${\mathbf r}_0$ and ${\mathbf r}'$ can easily be performed, and the result is consistent with that
of Ref.\ \cite{ref:Cucchietti04}.

Eq.\ \eq{eq4} still involves rapidly oscillating phases, and again most of the contributions will cancel out, 
unless the trajectories in a pair are systematically correlated. The main contribution corresponds to the 
diagonal approximation, i.e. $\gamma=\gamma '$, which gives the classical survival probability. Together with the sum rule \cite{ref:Sieber99} for open systems, this yields 
\beq\label{cdec}
\bar{\rho}^{\rm diag}(t) =\langle {\rm e}^{-t/ \tau_d}\rangle_{{\mathbf r},{\mathbf p}} \, ,
\eeq
where $\langle ...\rangle_{{\mathbf r},{\mathbf p}}$ indicates a phase space average, 
\beq
\langle  F \rangle_{{\mathbf r},{\mathbf p}}=\frac{1}{(2\pi \hbar)^2}\int d{\mathbf r}d{\mathbf p} \,
F({\mathbf r},{\mathbf p})\rho_{W}({\mathbf r},{\mathbf p}) \, ,
\eeq  
and $1/\tau_d$ is the classical escape rate at the energy $E=H({\mathbf r},{\mathbf p})$, where $H({\mathbf r},{\mathbf p})$ is the Hamiltonian of the system. 
For a two-dimensional system, $\tau_d=\Omega(E)/(2 w p)$, 
with $\Omega(E)=\int d{\mathbf r}'d{\mathbf p}'\delta(E-H({\mathbf r}',{\mathbf p}'))$, $w$ 
the size of the opening, and $p=|{\mathbf p}|$. For a chaotic billiard this reduces to 
$\tau_d=m\pi A/(w p)$. For an initial state with a well defined energy $E_0$ we can write
$\bar{\rho}^{\rm diag}(t) ={\rm e}^{-t/ \tau_d(E_0)}$. 
In the following we will assume this to be the case and drop the brackets of the phase space 
average.

Equation \eq{cdec} has two restrictions: First, we have supposed that at time $t$ the trajectories can 
already be considered ergodic (they have homogeneously explored the phase space). This is a good 
assumption as long as  $t\lambda\gg 1$, with $\lambda$ being the Lyapunov exponent. 
Second, we have assumed that the ergodicity of the corresponding closed system is not affected 
by the opening, meaning, classically the opening should be small $\tau_d\lambda\gg 1$, while
quantum mechanically it is very large $\tau_d\ll t_H$.


\section{III. Survival probability:
           Leading order weak localization-type contributions}

\bfg
\centerline{ \includegraphics[width=8cm]{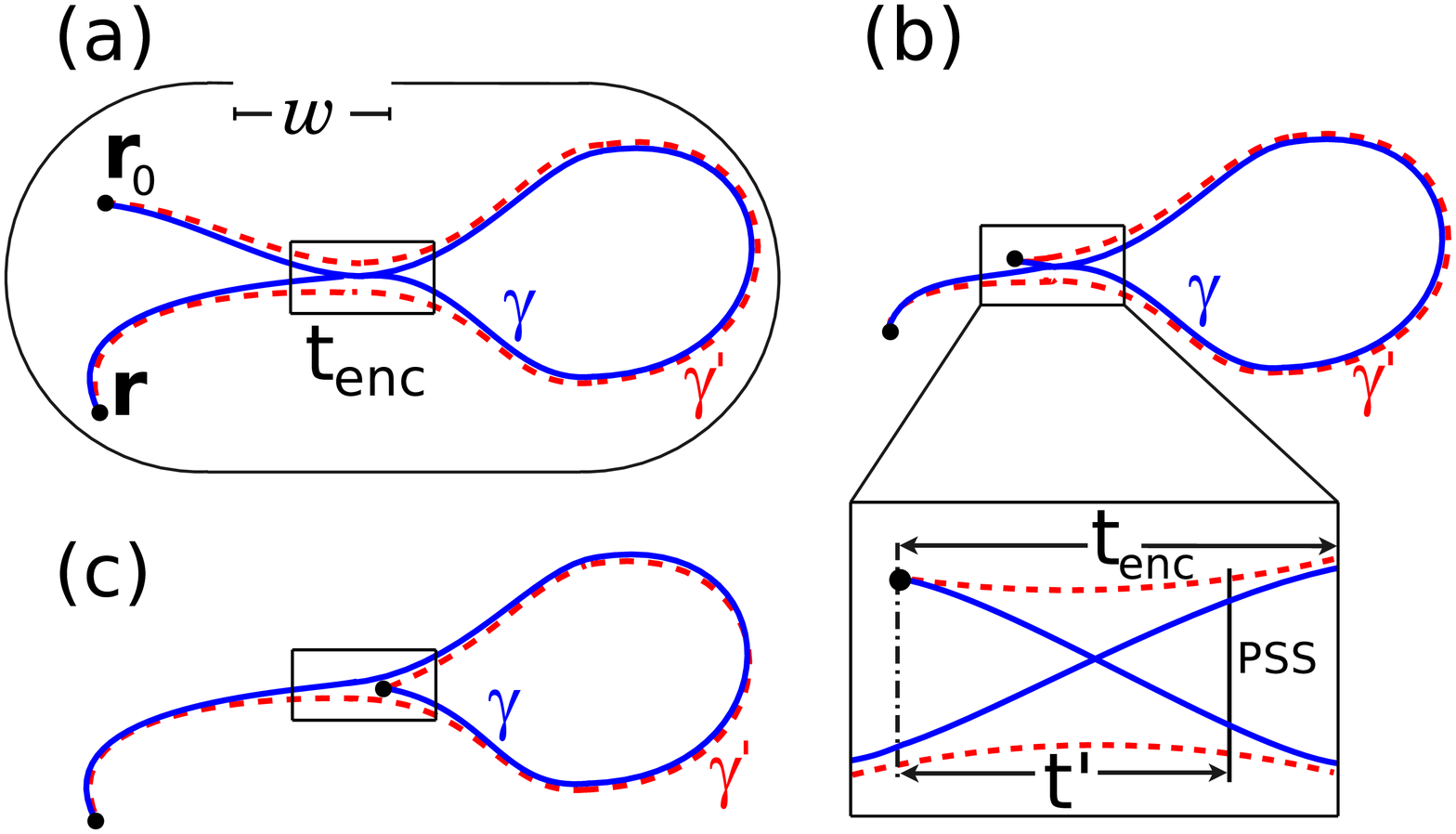}}
\caption{(color online) Scheme of two-leg-loop (2ll, a) and one-leg-loop (1ll, b,c) orbit pairs.
The trajectories $\gamma$ (full line) and $\gamma'$ (dashed) connect the points ${\bf r}_0$ with ${\bf r}$ in a time $t$, and they differ by a 2-encounter in (a). When the beginning or the end of the trajectory is inside the encounter we have the situation plotted in (b). (c) is a variation of (b) where there is no self-crossing of either of the two trajectories. }\label{fig:loops1}
\efg

It was shown in Ref.\ \cite{ref:Waltner08} that the leading quantum corrections to the semiclassical
survival probability \eq{smcsp} for systems with time reversal symmetry come from orbits with a self 
encounter (\fig{fig:loops1}a), `two-leg-loops' (2ll or 2-encounter) introduced in Ref.\
\cite{ref:Sieber01}, together with `one-leg-loops' (1ll, sketched in \fig{fig:loops1}b,\ c), 
which together preserve unitarity.

\subsection{Two-leg-loops}

In this section we will give a detailed derivation of these contributions to the survival probability 
following the phase space approach \cite{ref:Heusler04}. The double sum over trajectories is replaced 
by the sum rule together with integrals over the stable and unstable manifolds along reference
trajectories $\gamma$ weighted by the density of 2-encounters in a orbit of length $t$, 
$w^{\rm 2ll}(u,s,t)$, giving rise to a difference in action $\Delta S(u,s)=us$, whose absolute value is smaller than a classical 
value $c^2$. This density is given by 
\beq
w^{\rm 2ll}(u,s,t)=\frac{(t-2t_{\rm enc})^2}{2\Omega t_{\rm enc}},
\eeq
where the encounter time is $t_{\rm enc}=\lambda^{-1}\ln (c^2/|us|)$.

The  classical survival probability is modified by a factor ${\rm e}^{t_{\rm enc}/\tau_d}$, since the fact that the first stretch remains inside the cavity implies that the second will also be inside. Thus
\beq\label{s2ll}
\bar{\rho}^{\rm 2ll}(t)=
{\rm e}^{-t/ \tau_d} \int_{-c}^{c}\!\! du\!\int_{-c}^{c}\!\! ds \, w^{\rm 2ll}(u,s,t)e^{t_{\rm enc}/ \tau_d} {\rm e}^{\frac{i}{\hbar}us}. 
\eeq
The integration can be performed by making the change of variables $x=us/c^2$, $\sigma=c/u$ as in Ref.\ \cite{ref:Brouwer06}. The result is
\beq
\label{rho-2ll}
\bar{\rho}^{\rm 2ll} (t) =  e^{-t/ \tau_d}\left(\frac{t^2}{2\tau_d t_H}-2\frac{t}{t_H}\right).
\eeq
The quadratic term corresponds to the first order quantum correction according to Ref.\ \cite{ref:Frahm97}, while the linear term breaks unitarity, since it does not vanish as $\tau_d\to \infty$ (when the system is closed). As shown in Ref.\ \cite{ref:Waltner08} another type of diagram has to be considered in order to solve this problem.

\subsection{One-leg-loops}
The relevant diagrams correspond to trajectories with an encounter at the 
beginning or at the end of the trajectory, as shown in \fig{fig:loops1}b,\ c. Clearly, the latter only 
exists for initial and final points inside the cavity, since at the openings the exit of one stretch of 
the encounter implies the exit of the other one (with perfect coupling).  

To evaluate these two contributions we define a Poincar\'e surface of section at some time $t'$ from the end 
or beginning of the trajectory \cite{ref:Brouwer06}. The encounter time will be given by
\beq
t_{\rm enc}(t',u)=t' +\frac{1}{\lambda}\ln(c/|u|)\, ,
\eeq
with the restriction $t'<\frac{1}{\lambda}\ln(c/|s|)$, while the density of such encounters is given by
\bea
w^{\rm 1ll}(u,s,t)&=&2\int_{0}^{\frac{1}{\lambda}\ln\frac{c}{|s|}} \! dt' \! \int_{0}^{t-2 t_{\rm enc}}dt_2 \frac{1}{\Omega t_{\rm enc}(t' ,u)} \nonumber\\
&=&2 \int_{0}^{\frac{1}{\lambda}\ln\frac{c}{|s|}} dt'  \frac{t-2t_{\rm enc}(t',u)}{\Omega t_{\rm enc}(t',u)}.
\eea
The factor two is due to the possibility of having the encounter at the beginning of the trajectory 
or at the end. The difference in action will be $\Delta S\approx us$ at any point of the Poincar\'e 
surface of section. It is important to mention that this weight function automatically includes the 
situation where both end points are very close i.e.\ coherent back-scattering. We can now proceed to 
calculate this contribution to the survival probability in the same way as before, replacing 
$w^{\rm 2ll}(u,s,t)$ by $w^{\rm 1ll}(u,s,t)$ in Eq.\ \eq{s2ll}.
In order to evaluate the integrals, we make the change of variables \cite{ref:Brouwer06}
\beq\label{chavar}
t''=t'+\frac{1}{\lambda}\ln\left(\frac{c}{|u|}\right),\qquad u=c/\sigma,\qquad s=cx\sigma,
\eeq
with an integration domain $-1<x<1$, $1<\sigma<e^{\lambda t''}$ and 
$0<t''< \frac{1}{\lambda}\ln\left(\frac{1}{|x|}\right) $. Here is important to notice that the limits 
of $t''$ also include the situation where the point at which the orbits start 
is after a possible self-crossing. This means that it is not necessary to have a true self-crossing 
in configuration space in order to give a contribution of this kind. 

We define $\bar{\rho}^{\rm 1ll}(t)=I{\rm e}^{-t/\tau_d}$ where,
\beq
I=2\int_{-c}^{c} du \int_{-c}^{c} ds \int_{0}^{\frac{1}{\lambda}\ln\frac{c}{|s|}} dt'  \frac{t-2t_{\rm
enc}}{\Omega t_{\rm enc}}  {\rm e}^{\frac{i}{\hbar} us}{\rm e}^{t_{\rm enc}/\tau_d} \, ,
\eeq
the integral over $\sigma$ can be easily done after the change of variables mentioned above,
and $I$ can be written as
\bea\label{eqIa}
I &=& \frac{4r\lambda}{\pi t_H}\int_0^1 dx \cos(r x)\int_{0}^{\frac{1}{\lambda}\ln(1/x) } dt''   (t-2t'')  {\rm e}^{t''/ \tau_d} \nonumber\\ 
&=& \left(t-2\frac{d}{d\tau_d^{-1}}\right)\frac{4r\lambda\tau_d}{\pi t_H}\int_0^1 dx \cos(rx)
x^{-\frac{1}{\lambda \tau_d}} \, ,
\eea
where $r=c^2/\hbar$.

The integration over $x$ can be performed by parts, neglecting highly oscillating terms that will disappear 
after averaging \cite{ref:Brouwer06}, yielding
\bea
I&=&\left(t-2\frac{d}{d\tau_d^{-1}}\right)\frac{4r \lambda\tau_d}{\pi t_H}\left(\frac{\sin(r)}{r}+\frac{1}{\lambda\tau_d}\int_0^1 dx \frac{\sin(r x)}{r x}\right)\nonumber\\ 
&=&\frac{4 t}{\pi t_H}\int_0^{r} dy\frac{\sin(y)}{y}\approx \frac{4 t}{\pi t_H}\int_0^{\infty} dy\frac{\sin(y)}{y}=\frac{2 t}{ t_H}.
\eea
Then the 1ll contribution to the decay reads
\beq
\bar{\rho}^{\rm 1ll}\left(t\right) = 2\frac{t}{t_H}e^{-t/ \tau_d} \, .
\eeq
This term exactly cancels the linear term in Eq.~\eq{rho-2ll} coming from the 2ll contribution, 
recovering unitarity. The leading semiclassical correction (quadratic in time) to the classical
survival probability is therefore \cite{ref:Waltner08}
\beq
\bar{\rho}^{\rm 2ll+1ll} = \frac{t^2}{2\tau_d t_H} e^{-t/ \tau_d},
\eeq
which is consistent with the RMT prediction \cite{ref:Frahm97}. It can be interpreted as
a interference-based  weak localization-type enhancement of the survival probability.

In the next section we will extend this approach to include higher order corrections, 
coming from semiclassical diagrams with multiple encounters or with one encounter involving multiple stretches. 


\section{IV. Survival probability: 
           Higher order contributions for the GUE, GOE and GSE cases}
\label{secHOC}
\bfg
\centerline{ \includegraphics[width=8cm]{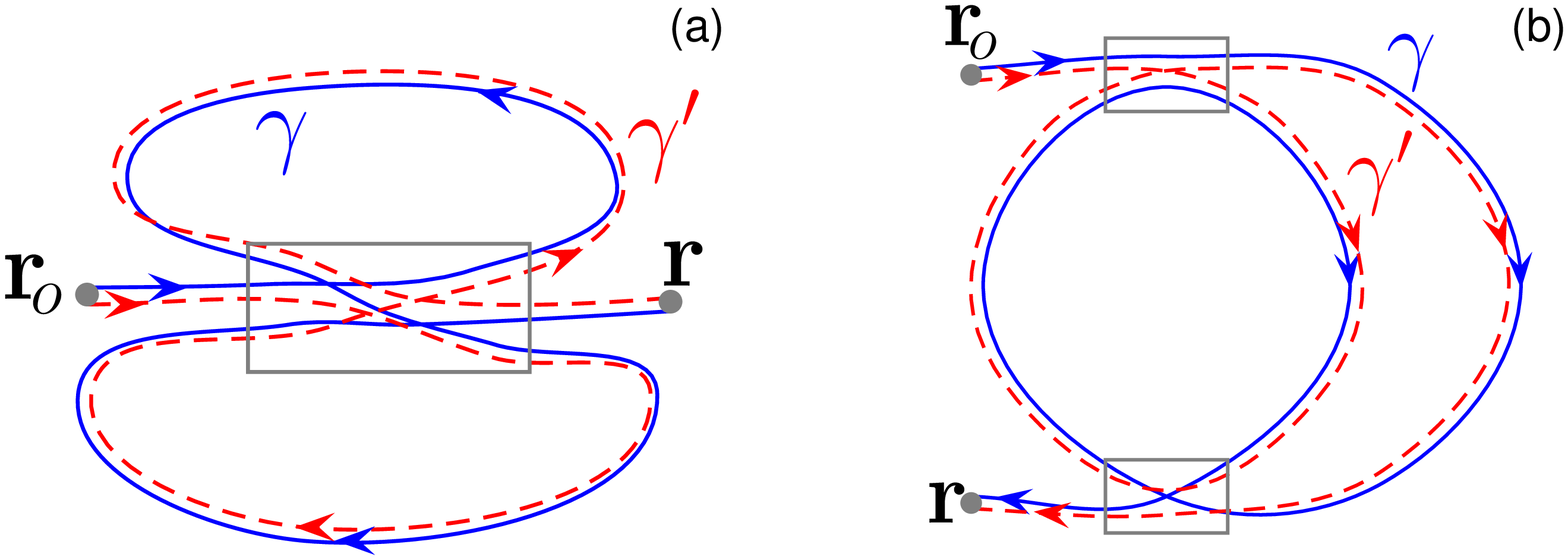}}
\caption{(color online) Scheme of orbit pairs that do not require time-reversal symmetry that give higher order corrections: (a) a single 3-encounter, (b) a double 2-encounter. The trajectories $\gamma$ (full line) and $\gamma'$ (dashed) connect the points ${\bf r}_0$ with ${\bf r}$ in a time $t$, and they differ by the way they are connected in the encounter regions. 
}\label{fig:loops2}
\efg
For the unitary case, the next order contributions to $\rho(t)$ are given by the diagrams shown
in \fig{fig:loops2}, as indicated in Ref.\ \cite{ref:Heusler04}. In a similar way, we can compute 
the next order corrections for systems with time reversal symmetry. Time reversal symmetry, however,
allows more structures, the corresponding diagrams include the ones sketched in \fig{fig:loops2} 
(multiplied by a factor of 4 for \fig{fig:loops2}a and a factor of 3 for \fig{fig:loops2}b, 
respectively \cite{ref:Heusler04}) together with a structure including two copies of the 
encounter in \fig{fig:loops1}a. 

In general, an encounter region contains an arbitrary number of $l\ge 2$ stretches of the trajectory, which are mutually linearizable, and one speaks of an $l$-encounter. In order to calculate higher order corrections, we consider trajectory pairs with encounters described by the vector $\v$, whose elements $v_l$ list the number of $l$-encounters in the trajectory pair.  The total number of encounters is then $V=\sum v_{l}$ while the number of links of the related closed orbit is $L=\sum lv_{l}$ as in Ref.\ \cite{ref:Heusler06}. 
 
Consider a periodic orbit formed by joining the ends of the open orbit. We can generate the open trajectories by cutting this closed orbit along each of its links and moving the ends of the cut to the required positions.  Note that for systems with time reversal symmetry, we must chose either the partner orbit or its time reversal so that the link, which is cut, is traversed in the same direction by both orbits.  The contribution can then be separated into three parts:
\begin{description}
\item[A] where the start and end points are outside of the encounters (2ll),
\item[B] where either the start or end point is inside an encounter (1ll) and 
\item[C] where both the start and end point are inside encounters (0ll).
\end{description}

\subsection{Case A}
This contribution can be written as
\bea \label{Icontribeqn}
\bar{\rho}_{\v, \mathrm{A}}(t)&=&N(\v)\int \ud\s\:\ud\u\:w_{\v,\mathrm{A}}(\u,\s,t)\rme^{-\mu t} 
\nonumber \\
&\times& \rme^{\sum_{\alpha=1}^{V}(l_{\alpha}-1)\mu t_{\mathrm{\rm enc}}^{\alpha}}\rme^{\frac{\ci}{\hbar}\u\s},
\eea
where $N(\v)$ is the number of trajectory structures corresponding to each vector $\v$ \cite{ref:Heusler04}, $\mu=1/\tau_d$, and $\alpha$ labels the $V$ encounters, each being an $l_{\alpha}$-\ encounter. We have included the correction to the survival probability of the trajectories due to the proximity of encounter stretches during the encounters.  In terms of an integral the weight is given by
\beq \label{weightinttrajeqn}
w_{\v,\mathrm{A}}(\u,\s,t)=\frac{\int_{0}^{t-t_{\mathrm{\rm enc}}}\ud t_{L}\ldots\int_{0}^{t-t_{\mathrm{\rm enc}}-t_{L}\ldots-t_{2}}\ud t_{1}}{\Omega^{L-V}\prod_{\alpha}t_{\mathrm{\rm enc}}^{\alpha}},
\eeq
where $t_{\mathrm{\rm enc}}$ is the total time that the trajectory spends in the encounters $t_{\mathrm{\rm enc}}=\sum_{\alpha=1}^{V}l_{\alpha}t_{\mathrm{\rm enc}}^{\alpha}$. Each of the links must have positive duration and this restriction is included in the limits of integration. The weight is simply an $L$-fold integral over different link times $t_{i}, i=1\ldots L$, while the last link time is fixed by the total trajectory time
\beq
t=\sum_{i=1}^{L+1}t_{i}+\sum_{\alpha=1}^{V}l_\alpha t_{\rm enc}^\alpha \, .
\eeq
When we perform the integrals the weight function becomes
\beq \label{weighttrajeqn}
w_{\v,\mathrm{A}}(\u,\s,t)=\frac{\left(t-\sum_{\alpha}l_{\alpha}t_{\mathrm{\rm enc}}^{\alpha}\right)^L}{L!\Omega^{L-V}\prod_{\alpha}t_{\mathrm{\rm enc}}^{\alpha}}.
\eeq
To calculate the semiclassical contribution we will rewrite Eq.\ \eq{Icontribeqn} as
\beq
\bar{\rho}_{\v, \mathrm{A}}(t)=N(\v)\int \ud\s\:\ud\u\:z_{\v,\mathrm{A}}(\u,\s,t)\rme^{-\mu t}\rme^{\frac{\ci}{\hbar}\u\s},
\eeq
where $z_{\v,\mathrm{A}}(\u,\s,t)$ is an augmented weight including the term from the survival probability correction of the encounters
\bea \label{zeqn}
z_{\v,\mathrm{A}}(\u,\s,t)&=&w_{\v,\mathrm{A}}(\u,\s,t)\rme^{\sum_{\alpha}(l_{\alpha}-1)\mu t_{\mathrm{\rm enc}}^{\alpha}}\\ \nonumber
&\approx &\frac{\left(t-\sum_{\alpha}l_{\alpha}t_{\mathrm{\rm enc}}^{\alpha}\right)^L\prod_{\alpha}\left(1+(l_{\alpha}-1)\mu t_{\mathrm{\rm enc}}^{\alpha}\right)}{L!\Omega^{L-V}\prod_{\alpha}t_{\mathrm{\rm enc}}^{\alpha}},
\eea
where we have expanded in the second line the exponent to first order in the encounter times.  We can now use the fact that the semiclassical contribution comes from terms where the encounter times in the numerator cancel those in the denominator exactly \cite{ref:Heusler04}.  Keeping only those terms, we then obtain a factor of $(2\pi\hbar)^{L-V}$ from the integrals over $\s$ and $\u$ and obtain the result for trajectories described by the vector $\v$ of interest.

Consider for example a trajectory with a 3-encounter with two long legs, sketched in \fig{fig:loops2}a. 
The encounter has a duration given by 
\beq
t_{\rm enc}\approx \frac{1}{\lambda} \ln \frac{c^2}{{\rm max}_j |s_j|\times {\rm max}_j |u_j|},
\eeq
where $j=1,2$ and $u_j$, $s_j$ are the differences between the unstable and stable coordinates of the trajectory on the PSS placed in the encounter region, respectively.

The density of this type of encounter, with an action difference $\Delta S=\u\cdot \s$, is
\beq
w_{(3)^1,{\mathrm A}}(\u,\s,t)= \frac{(t-3t_{\rm enc})^3}{6\Omega^2t_{\rm enc} },
\eeq
where we use the notation $(l)^{v_l}$ to indicate that the trajectory has $v_l$ $l$-encounters.
We can calculate the contribution of such orbits by replacing the sum over the partner trajectory
$\gamma'$ with an integral over the stable and unstable coordinates $(\u,\s)$ with the density
$w_{(3)^1,{\mathrm A}}(\u,\s,t)$, modifying the classical survival probability entering the sum rule by
a factor ${\rm e}^{2\mu t_{\rm enc}}$. In the case of time reversal symmetry there are four possible
structures in this case \cite{ref:Heusler06}, and the final result is
\beq\label{GUE-1}
\bar{\rho}_{(3)^1,{\mathrm A}}(t) = 4 e^{-t/ \tau_d}\left(-\frac{3t^2}{2 t_H^2}+\frac{t^3}{3 \tau_d t_H^2} \right).
\eeq
For a double 2-encounter shown in \fig{fig:loops2}b, we define two encounter times: $t_{\rm enc}^{1} 
\approx \frac{1}{\lambda} \ln \frac{c^2}{|u_1 s_1|} $ and $t_{\rm enc}^{2} 
\approx \frac{1}{\lambda}\ln \frac{c^2}{|u_2 s_2|} $. 
\\The density of such a double-encounter is given by
\beq
w_{(2)^2,{\mathrm A}}(\u,\s,t)= \frac{(t-2t_{\rm enc})^4}{24\Omega^2t_{\rm enc}^{1}t_{\rm enc}^{2} },
\eeq
with $t_{\rm enc}= t_{\rm enc}^{1}+t_{\rm enc}^{2}$.
In this case the number of possible structures for systems with time reversal symmetry is 5. 
The contribution of such orbits to the survival probability is 
\beq\label{GUE-2}
\bar{\rho}_{(2)^2,{\mathrm A}}(t) = 5 e^{-t/ \tau_d}\left(2\frac{t^2}{t_H^2}-\frac{2t^3}{3 \tau_d t_H^2}+\frac{t^4}{24 \tau_d^2 t_H^2} \right).
\eeq

The total contribution of structures with $L-V=2$ of 2ll's is then
 \beq\label{GOE2llho}
\bar{\rho}_{2,{\mathrm A}}(t) = e^{-t/ \tau_d}\left(4\frac{t^2}{t_H^2}-\frac{2t^3}{\tau_d t_H^2}+\frac{5t^4}{24 \tau_d^2 t_H^2} \right).
\eeq

\subsection{Case B}
Now we have to consider the corresponding `one-leg-loops' for the previous diagrams.
This contribution can be written as
\beq \label{IIcontribeqn}
\bar{\rho}_{\v, \mathrm{B}}(t)=N(\v)\int \ud\s\:\ud\u\:z_{\v,\mathrm{B}}(\u,\s,t)\rme^{-\mu t}\rme^{\frac{\ci}{\hbar}\u\s}.
\eeq
Here one encounter overlaps with the start or end of the trajectory, we have therefore one link fewer
($L$ in total) and an extra integral over the position of the encounter relative to the starting point.
Starting with a closed periodic orbit, (and dividing by the overcounting factor of $L$) we can cut each
of the $L$ links in turn and move the encounter on either side of the cut to either the start or the
end.  In total we obtain $l_{\alpha'}$ copies of the same 1ll involving the encounter $\alpha'$, and
additional factor of 2 appears due to the possibilities of having the encounter at the beginning or at
the end of the trajectory.  The augmented weight can then be expressed as a sum over the different possibilities,
each of which involves an integral over the distance from the PSS to the initial or final point,
$t_{\alpha'}$,
\bea \label{weighttrajeqnII}
z_{\v,\mathrm{B}}(\u,\s,t)&=&2\sum_{\alpha'=1}^{V}l_{\alpha'}\int \ud t_{\alpha'}\frac{\left(t-\sum_{\alpha}l_{\alpha}t_{\mathrm{\rm enc}}^{\alpha}\right)^{L-1}}{L!\Omega^{L-V}\prod_{\alpha}t_{\mathrm{\rm enc}}^{\alpha}}\nonumber \\ &&\times \rme^{\sum_{\alpha=1}^{V}(l_{\alpha}-1)\mu t_{\mathrm{\rm enc}}^{\alpha}}.
\eea
Because of the integrals over the position of the encounter at the start or end of the trajectory, the semiclassical contribution is calculated differently, using integrals of the type we encountered in Eq.\ \eq{eqIa}. However, it is easy to see in Eq.\ \eq{eqIa} that after a suitable change of variables, the integral over $\tau$ can be effectively replaced by a $t_{\rm enc}$. This change of variables can be done for each $(u_{\alpha'},s_{\alpha'},t_{\alpha'})$, giving again a factor of  $t_{\rm enc}^{\alpha'}$ for each integral over $t_{\alpha'}$, so that the augmented weight can be written as
\bea \label{zeqnII}
z_{\v,\mathrm{B}}(\u,\s,t)&\approx&\frac{2\left(\sum_{\alpha}l_{\alpha}t_{\mathrm{\rm enc}}^{\alpha}\right)\left(t-\sum_{\alpha}l_{\alpha}t_{\mathrm{\rm enc}}^{\alpha}\right)^{L-1}}{L!\Omega^{L-V}\prod_{\alpha}t_{\mathrm{\rm enc}}^{\alpha}}\nonumber\\ &\times& \prod_{\alpha}\left(1+(l_{\alpha}-1)\mu t_{\mathrm{\rm enc}}^{\alpha}\right),
\eea
and treated as before.

For a single 3-encounter, we define again a Poincar\'e surface of section at a time $t'$ from the
beginning or end of the orbit. The encounter time is given by 
$t_{\rm enc}(t',u_{\rm max})=t'+\frac{1}{\lambda}\ln (c/{\rm max_i}\, |u_i|)$, 
and the augmented weight of such encounter
\beq\label{w311ll}
z_{(3)^1,\mathrm{B}} (\u,\s,t)=\int_0^{\frac{1}{\lambda}\ln \frac{c}{|s|_{\rm max}}}dt' \frac{(t-3t_{\rm enc})^2}{\Omega^2t_{\rm enc} }\rme^{2\mu t_{\rm enc}}. 
\eeq
Making the change of variables indicated in Eq.\ \eq{chavar} and multiplying by the number of possible structures, the resulting contribution for systems with time reversal symmetry is
\beq\label{GUE-3}
\bar{\rho}_{(3)^1,\mathrm{B}}(t) =  4e^{-t/ \tau_d}\left(\frac{t^2}{t_H^2}\right).
\eeq
The integration of Eq.\ \eq{w311ll} yields the same result as if we had used instead the augmented weight function $ \frac{(t-3t_{\rm enc})^2}{\Omega^2}(1+2\mu t_{\rm enc})$, as in Eq.\ \eq{zeqnII}.

For the double 2-encounter one of the Poincar\'e surfaces of sections will be from the beginning 
(or end) of the trajectory at a time $t'$, then $t_{\rm enc}^1(t',u_1)=t'+ \frac{1}{\lambda}\ln |c/ u_1|$, and the corresponding density of such pairs is
\beq
z_{(2)^2,\mathrm{B}}(\u,\s,t) =  
\frac{1}{3}\int_0^{\frac{1}{\lambda}\ln \frac{c}{|s_1|}}dt' \frac{(t-2t_{\rm enc})^3}{\Omega^2 t_{\rm enc}^1t_{\rm enc}^2}\rme^{\mu t_{\rm enc}}.
\eeq
We perform the same change of variables as before for $(u_1,s_1,\tau)$, which yields
\beq\label{GUE-4}
\bar{\rho}_{(2)^2,\mathrm{B}}(t) =  5e^{-t/ \tau_d}\left(\frac{t^3}{3t_H^2\tau_d}-2\frac{t^2}{t_H^2} \right).
\eeq
The total contribution of 1ll's for $L-V=2$ for systems with time reversal symmetry is given by
 \beq\label{GOE1llho}
\bar{\rho}_{2,\mathrm{B}}(t) = e^{-t/ \tau_d}\left(\frac{5t^3}{3t_H^2\tau_d}-6\frac{t^2}{t_H^2} \right).
\eeq

\subsection{Case C}

This contribution can be written as
\beq \label{IIIcontribeqn}
\bar{\rho}_{\v, \mathrm{C}}(t)=\int \ud\s\:\ud\u\:z_{\v,\mathrm{C}}(\u,\s,t)\rme^{-\mu t}\rme^{\frac{\ci}{\hbar}\u\s}.
\eeq

Now that we have one encounter overlapping with the start of the trajectory, and a second (different) encounter at the end of the trajectory, we have several additional complications.  Firstly, there is again one link fewer ($L-1$ in total) and now we have two extra integrals over the position of the start and end encounters relative to the start and end point.  Also the number of such structures is different.  Starting with a closed periodic orbit, we can cut each of the $L$ links in turn and move the encounters on either side of the cut to both the start and the end, as long as the link joins two different encounters.  We therefore need to count the number of ways that this is possible for the different sizes of encounters that are linked.  We record these numbers in a matrix $\mathcal{N}(\v)$, where the elements $\mathcal{N}_{\alpha,\beta}(\v)$ record the number of links (divided by $L$) linking encounter $\alpha$ with encounter $\beta$, in this case it is convenient to include $\mathcal{N}_{\alpha,\beta}(\v)$ in the augmented weight function. The augmented weight, including these possibilities, can then be expressed as the following sum over the 0ll encounters 
\bea \label{weighttrajeqnIII}
z_{\v,\mathrm{C}}(\u,\s,t)&=&\sum_{\alpha',\beta'}\mathcal{N}_{\alpha',\beta'}(\v)\int \ud t_{\alpha'}\ud t_{\beta'}\rme^{\sum_{\alpha=1}^{V}(l_{\alpha}-1)\mu t_{\mathrm{\rm enc}}^{\alpha}}\nonumber\\ &&\times\frac{\left(t-\sum_{\alpha}l_{\alpha}t_{\mathrm{\rm enc}}^{\alpha}\right)^{L-2}}{(L-2)!\Omega^{L-V}\prod_{\alpha}t_{\mathrm{\rm enc}}^{\alpha}} .
\eea
Again we can expand the exponent to first order in the encounter times and write the augmented weight function as 
\bea \label{zeqnIII}
z_{\v,\mathrm{C}}(\u,\s,t)&\approx&\left(\sum_{\alpha,\beta}\mathcal{N}_{\alpha,\beta}(\v)t_{\mathrm{\rm enc}}^{\alpha}t_{\mathrm{\rm enc}}^{\beta}\right)\left(t-\!\sum_{\alpha}l_{\alpha}t_{\mathrm{\rm enc}}^{\alpha}\right)^{L\!-\!2} \nonumber\\
&& \times \frac{\prod_{\alpha}\left(1+(l_{\alpha}-1)\mu t_{\mathrm{\rm enc}}^{\alpha}\right)}{(L-2)!\Omega^{L-V}\prod_{\alpha}t_{\mathrm{\rm enc}}^{\alpha}},
\eea
and treating it as before.\\

For a single 3-encounter there cannot be such a contribution. For a double 2-encounter we may define two Poincar\'e surfaces of section at $t'_1$ and $t'_2$ from the beginning and the end of the trajectory, $t_{\rm enc}^1(t'_1,u_1)=t'_1+ \frac{1}{\lambda}\ln |c/ u_1|$ and $t_{\rm enc}^2(t'_2,u_2)=t'_2+ \frac{1}{\lambda}\ln |c/ u_2|$, and the corresponding density 
\bea
z_{(2)^2,\mathrm{C}}(\u,\s,t)&=&\int_0^{\frac{1}{\lambda}\ln \frac{c}{|s_1|}}\!\!dt'_1\int_0^{\frac{1}{\lambda}\ln\frac{c}{|s_2|}}\!\!dt'_2 \rme^{\mu t_{\rm enc} }\nonumber\\ && \times \frac{(t-2t_{\rm enc})^2}{2\Omega^2 t_{\rm enc}^1t_{\rm enc}^2}.
\eea
This gives the following contribution to the survival probability
\beq\label{GUE-5}
\bar{\rho}_{(2)^2,\mathrm{C}}(t) =  e^{-t/ \tau_d}\left(\frac{2t^2}{t_H^2} \right) \, .
\eeq
Again the result is the same as if we had used 
$z_{(2)^2,\mathrm{C}}(\u,\s,t)=2\frac{(t-2t_{\rm enc})^2}{\Omega^2}(1+\mu t_{\rm enc})$, 
which corresponds to Eq.\ \eq{zeqnIII}.

\subsection{Unitary case}

We can easily calculate the contribution for each vector $\v$ for each of the three cases, 
as long as we know the numbers of possible trajectory structures. For cases A and B, 
these numbers can be found in Ref.\ \cite{ref:Heusler04} and are repeated in the first four 
columns of Table \ref{guenumbers}. For case C we will go up to the sixth order correction, 
$L-V=6$, and for this we have at most three different types of $l$-encounters. It is useful 
to rewrite the sum over $\alpha$ and $\beta$ as a sum over the components of the vector $\v$. $\mathcal{N}_{\alpha,\beta}(\v)$ records the number of ways of 
cutting links that connect encounter $\alpha$ and $\beta$, in the periodic orbit structures 
described by $\v$. However we can see that the important quantities are the sizes of the 
encounter $\alpha$ and $\beta$. Instead we record in $\mathcal{N}_{k,l}(\v)$ the number of links that join 
an encounter of size $k$ to an encounter of size $l$. If we number the encounters from 1 to $V$ in 
order of their size, then we only need to know the numbers $\mathcal{N}_{l_1,l_2}(\v)$, 
$\mathcal{N}_{l_1,l_V}(\v)$ and $\mathcal{N}_{l_{V-1},l_V}(\v)$, as the maximal number of different sized encounters is three. Moreover $\mathcal{N}_{k,l}$ is symmetric, therefore we include in Table \ref{guenumbers} both $\mathcal{N}_{k,l}$ and $\mathcal{N}_{l,k}$ together. Using a program to count and classify the possible permutation matrices we 
obtain the remaining columns in Table \ref{guenumbers} for systems without time reversal symmetry. 
Note that certain encounter combinations might correspond to several elements of the numbers 
$\mathcal{N}_{l_1,l_2}(\v)$, $\mathcal{N}_{l_1,l_V}(\v)$ and $\mathcal{N}_{l_{V-1},l_V}(\v)$, 
in which case we record their number in the leftmost column.

\begin{table}[htb]
\centering
\begin{tabular}{|c|c|c|c|c|c|c|}
\hline
$\v$&$L$&$V$&$N(\v)$&$\mathcal{N}_{l_1,l_2}(\v)$&$\mathcal{N}_{l_1,l_V}(\v)$&$\mathcal{N}_{l_{V-1},l_V}(\v)$\\
\hline
$(2)^{2}$&4&2&1&1&&\\
$(3)^{1}$&3&1&1&&&\\
\hline
$(2)^{4}$&8&4&21&21&&\\
$(2)^{2}(3)^{1}$&7&3&49&12&32&\\
$(2)^{1}(4)^{1}$&6&2&24&16&&\\
$(3)^{2}$&6&2&12&8&&\\
$(5)^{1}$&5&1&8&&&\\
\hline
$(2)^{6}$&12&6&1485&1485&&\\
$(2)^{4}(3)^{1}$&11&5&5445&2664&2592&\\
$(2)^{3}(4)^{1}$&10&4&3240&984&1920&\\
$(2)^{2}(3)^{2}$&10&4&4440&464&2624&960\\
$(2)^{2}(5)^{1}$&9&3&1728&228&1080&\\
$(2)^{1}(3)^{1}(4)^{1}$&9&3&2952&552&760&1080\\
$(3)^{3}$&9&3&464&380&&\\
$(2)^{1}(6)^{1}$&8&2&720&360&&\\
$(3)^{1}(5)^{1}$&8&2&608&360&&\\
$(4)^{2}$&8&2&276&180&&\\
$(7)^{1}$&7&1&180&&&\\
\hline
\end{tabular}
\caption{The number of trajectory pairs and the number linking certain encounters for systems without time reversal symmetry.}
\label{guenumbers}
\end{table}

Table \ref{guenumbers} allows us to obtain the following results for the 
quantum corrections to the classical decay for the unitary case
\bea
\bar\rho_{2}(t) &=& \frac{\rme^{-\frac{t}{\tau_d}}}{t_H^2}\left(\frac{t^{4}}{24\tau_d^{2}}\right),\\
\bar\rho_{4}(t) &=& \frac{\rme^{-\frac{t}{\tau_d}}}{t_H^4}\left(\frac{t^{6}}{90\tau_d^{2}}-\frac{t^{7}}{180\tau_d^{3}}+\frac{t^{8}}{1920\tau_d^{4}}\right),\\
\bar\rho_{6}(t) &=& \frac{\rme^{-\frac{t}{\tau_d}}}{t_H^6}\left(\frac{t^{8}}{224\tau_d^{2}}-\frac{89t^{9}}{22680\tau_d^{3}}+\frac{31t^{10}}{30240\tau_d^{4}}\right. \nonumber \\
&& \qquad\qquad\left. -\frac{t^{11}}{10080\tau_d^{5}}+\frac{t^{12}}{322560\tau_d^{6}}\right).
\eea

These results enable us to calculate the decay up to 8th order in $t$, giving as the final result

\bea\label{GUE8or}
\bar{\rho}^{\mathrm{GUE}}(t) &=& \rme^{-\frac{t}{\tau_d}}\left(1+\frac{t^{4}}{24\tau_d^{2}t_H^2}+\frac{t^{6}}{90\tau_d^{2}t_H^4}-\frac{t^{7}}{180\tau_d^{3}t_H^4}\right.\nonumber\\ &&\left.+\left(\frac{1}{1920\tau_d^4 t_H^4}+\frac{1}{224\tau_d^2 t_H^6}\right)t^8+\ldots\right).
\eea
\subsection{Orthogonal case}

Similarly, we can find all possible permutation matrices and obtain Table \ref{goenumbers} (see Appendix A) 
for systems with time reversal symmetry. This gives us the result up to 7th order in $t$
\bea\label{GOE7or}
\bar{\rho}^{\mathrm{GOE}}(t) &=& \rme^{-\frac{t}{\tau_d}}\left[1+\frac{t^{2}}{2\tau_dt_H}-\frac{t^{3}}{3\tau_dt_H^2}\right.\\
&&+\left(\frac{5}{24\tau_d^{2}t_H^2}+\frac{1}{3\tau_dt_H^3}\right)t^{4} \nonumber\\
&& 
-\left(\frac{11}{30\tau_d^2t_H^3}+\frac{2}{5\tau_dt_H^4}\right)t^{5}\nonumber \\
&&
+\left(\frac{41}{720\tau_d^3t_H^3}+\frac{7}{12\tau_d^{2}t_H^4}+\frac{8}{15\tau_dt_H^5}\right)t^{6}\nonumber \\
&& \left.-\left(\frac{29}{168\tau_d^{3}t_H^4}+\frac{14}{15\tau_d^{2}t_H^5}+\frac{16}{21\tau_d t_H^6}\right)t^{7} +\ldots\right].\nonumber
\eea
The predictions for the decay using supersymmetry techniques can be found in Ref.\ \cite{ref:Savin97}, 
where the integrals appearing there can be expanded in powers of $t/t_H$, following the steps indicated 
in Ref.\ \cite{ref:Kuipers07}. The results of these expansions agree with Eqs.\ \eq{GUE8or} and \eq{GOE7or}. 

\subsection{Spin-orbit interaction and the symplectic case}

Along with the cases with and without time reversal symmetry, there has recently been interest in a
semiclassical treatment corresponding to the symplectic RMT ensemble in different contexts, such as 
in spectral statistics \cite{ref:Heusler04} and in the quantum transmission through mesoscopic 
conductors in the Landauer-B\"uttiker approach \cite{ref:Zaitsev05,ref:Bolte07}. There the symplectic 
case is obtained by including in the Hamiltonian a classically weak spin-orbit interaction. 

In the following we study the effect of spin-orbit interaction on the survival probability.
The spin-orbit interaction is accounted for by replacing the Hamiltonian for the orbital dynamics,
$\hat H_0$ considered up to now, by  
\begin{equation} 
\hat H=\hat H_0+\hat{\mathbf s}\cdot \mathbf C \left(\hat{\mathbf x},\hat{\mathbf p}\right),
\end{equation}
with $\mathbf C \left(\hat{\mathbf x},\hat{\mathbf p}\right)$ characterizing the coupling of the
translational degrees of freedom to the spin operator $\hat {\mathbf s}$.

For weak spin-orbit interaction, the semiclassical propagator is similar to Eq.\ \eq{smcprop}, where the classical trajectories are the same as for the case without interaction \cite{ref:Bolte99}. The only modification appears in the prefactor $D_\gamma$ that contains now the additional factor $B_\gamma\left({\mathbf x}',{\mathbf p}',t\right)$, which is the   spin-$s$ representation of the spin propagator matrix 
$b_\gamma\left({\mathbf x}',{\mathbf p}',t\right)$, defined as the solution of \cite{ref:Bolte99},
\begin{equation}
\frac{d}{dt}b_\gamma\left({\mathbf x}',{\mathbf p}',t\right) +\frac{i}{2}{\pmb{\sigma}}\cdot{\mathbf C} \left({\mathbf X}(t),{\mathbf P}(t)
\right)  b_\gamma\left({\mathbf x}',{\mathbf p}',t\right) =0 \, ,
\end{equation}
with the initial condition $b_\gamma\left({\mathbf x}',{\mathbf p}',0\right)=1$. 
This propagator can be used now in the derivation of a modified formula for the survival probability 
in the case of spin orbit interaction. After replacing the initial state $|\psi_0\rangle$ introduced 
in Eq.\ \eq{tdeps} by $|\Psi_0\rangle \equiv\left|\left.\psi_0\otimes {\mathbf s}_0\right\rangle\right.$ 
containing additionally the initial spin state $|{\mathbf s}_0\rangle$, we obtain the matrix element $\left\langle {\mathbf s}_0\left| 
B_\gamma B_\gamma^\dagger\right|{\mathbf s}_0\right\rangle$ as an additional factor inside the double sum in Eq.\ \eq{eq4}. We are interested in the average 
behaviour of this quantity. Therefore, we analyse $\frac{1}{\left(2s+1\right)}{\rm Tr }(
B_\gamma B_\gamma^\dagger)$ with ${\rm Tr}$ denoting the trace in the spin space. 
This quantity was already considered in Ref.\ \cite{ref:Bolte07}, where by assuming the mixing property 
of the combined spin and orbital dynamics, {\em i.e.}\ full spin relaxation,
it was shown that we can effectively write
\begin{equation}
\label{eq:Tr}
\frac{1}{\left(2s+1\right) }{\rm Tr}\left(B_\gamma B_\gamma^\dagger\right)=\left(\frac{\left(-1\right)^{2s}}{2s+1}\right)^{L-V}
\end{equation}
with $L$ and $V$ defined as before. It is important to notice that the contribution from spin orbit interaction depends, apart from on
the spin quantum number $s$, only on the difference $L-V$. 
The term \eq{eq:Tr} can now be inserted as prefactor into the expressions in appendix A for the GOE case after choosing 
in each term the correct value of $L-V$. For 
$s=1/2$ this yields
\bea\label{GSE7or}
\bar{\rho}^{\mathrm{GSE}}(t) &=& \rme^{-\frac{t}{\tau_d}}\left[1-\frac{t^{2}}{4\tau_dt_H}-\frac{t^{3}}{12\tau_dt_H^2}\right.\\
&&+\left(\frac{5}{96\tau_d^{2}t_H^2}-\frac{1}{24\tau_dt_H^3}\right)t^{4} \nonumber\\
&& 
+\left(\frac{11}{240\tau_d^2t_H^3}-\frac{1}{40\tau_dt_H^4}\right)t^{5}\nonumber \\
&& 
-\left(\frac{41}{5760\tau_d^3t_H^3}-\frac{7}{192\tau_d^{2}t_H^4}+\frac{1}{60\tau_dt_H^5}\right)t^{6}\nonumber \\ \nonumber
&& \left.-\left(\frac{29}{2688\tau_d^{3}t_H^4}-\frac{7}{240\tau_d^{2}t_H^5}+\frac{1}{84\tau_d
t_H^6}\right)t^{7}\right] \, .
\eea
This result is again consistent with RMT-type results for the symplectic ensemble
 \cite{ref:Frahm97}. The second, negative term in \eq{GSE7or} reflects weak-antilocalization
 effects in the quantum decay.


\section{V. Variance of the Decay}

In Sec.\ II we introduced a local time average, in order to select from the trajectories contributing
to the double sum in Eq.~(\ref{smcsp}) those that start from the same point. In order to compare 
deviations of $\rho(t)$ from the time-averaged $\bar{\rho}(t)$, we consider on the level of the 
diagonal approximation, the variance of $\bar{\rho}(t)$, averaged again over a time window:
\beq\label{var1}
\delta\bar{\rho}^2(t)=\left\langle (\rho(t)-\bar{\rho}(t))^2\right\rangle_{\Delta t}.
\eeq 
Substituting Eq.\ \eq{smcprop} in Eq.\ \eq{var1}, we can write the variance as
\bea\label{var2}
\delta\bar{\rho}^2(t) &=&\frac{1}{(2\pi \hbar)^{4}} {\Big \langle}
\int \prod_{i=1}^6 d{\mathbf r}_i \psi_0({\mathbf r}_1)\psi_0^*({\mathbf r}_2)\psi_0({\mathbf r}_4)\psi_0^*({\mathbf r}_5)\nonumber\\
&&
\!\times\sum_{\tilde\gamma_1 \left(\mathbf{r}_1\to\mathbf{r}_3,t\right)\atop \tilde\gamma_2
\left(\mathbf{r}_2\to\mathbf{r}_3,t\right) }\sum_{\tilde\gamma_3
\left(\mathbf{r}_4\to\mathbf{r}_6,t\right)\atop \tilde\gamma_4
\left(\mathbf{r}_5\to\mathbf{r}_6,t\right) }\tilde A {\rm e}^{\frac{i}{\hbar} 
\Delta S }{\Big\rangle}_{\Delta t} 
\eea
where $\tilde A=D_{\tilde\gamma_1 }D_{\tilde\gamma_2 }^* D_{\tilde\gamma_3}D_{\tilde\gamma_4}^*$ and
$ \Delta S= S_{\tilde\gamma_1} -S_{\tilde\gamma_2}+S_{\tilde\gamma_3}-S_{\tilde\gamma_4}$. Here
the configurations $\mathbf{r}_1\approx \mathbf{r}_2 $ and $\mathbf{r}_4\approx \mathbf{r}_5 $ have
already been taken into account in ${\bar\rho}(t)^2$ 
and therefore have to be ignored in Eq.\ \eq{var2}. Due to the average most of the contributions 
to Eq.\ \eq{var2} will cancel out, so for surviving the average the configuration of the points 
$\mathbf{r}_i$ must be such that the phase difference $\Delta {S}$ tends to zero. 
Apart from the configurations that contribute to $\bar{\rho}(t)$, the leading contribution comes 
from  $\mathbf{r}_1\approx \mathbf{r}_5 $ and $\mathbf{r}_2\approx \mathbf{r}_4 $, which 
requires $\mathbf{r}_3\approx \mathbf{r}_6$. We expand the trajectories $\tilde\gamma_1 $ and 
$\tilde\gamma_4 $ around trajectories $\gamma_1$ and $\gamma_4$ going from 
$\mathbf{q}_1=(\mathbf{r}_1+ \mathbf{r}_5)/2 $ to $\mathbf{q}_3=(\mathbf{r}_3+ \mathbf{r}_6)/2$ 
and  trajectories $\tilde\gamma_2 $ and $\tilde\gamma_3 $ around trajectories 
$\gamma_2$ and $\gamma_3$ going from $\mathbf{q}_2=(\mathbf{r}_2+ \mathbf{r}_4)/2 $ to $\mathbf{q}_3$. 
We can perform the integrals over $\mathbf{r}_1- \mathbf{r}_5 $ and 
$\mathbf{r}_2- \mathbf{r}_4$ and write the variance in terms of the Wigner function of the initial 
state, thus
\bea\label{var3}
  \delta\bar{\rho}^2(t)&=& \frac{1}{(2\pi \hbar)^{4}} {\Big \langle}
\int \prod_{i=1}^4 d{\mathbf q}_i \
\rho_W\left({\mathbf q}_1,\mathbf{\bar p}_{\gamma_1 \gamma_4}^0 \right) \\\nonumber
&&   \times \ \rho_W\left({\mathbf q}_2,\mathbf{\bar p}_{\gamma_2 \gamma_3}^0   \right)
\sum_{\gamma_1,\gamma_4 \left(\mathbf{q}_1\to\mathbf{q}_3,t\right)\atop \gamma_2,\gamma_3 \left(\mathbf{q}_2\to\mathbf{q}_3,t\right) }\!\!
\tilde A {\rm e}^{i\Delta S/\hbar}{\Big\rangle}_{\Delta t},\\ \nonumber
\eea
with
\beq
\mathbf{\bar p}_{\gamma_1 \gamma_4}^0= \frac{(\mathbf{p}_{\gamma_1,0}+\mathbf{p}_{\gamma_4,0})}{2} \, , \, 
\mathbf{\bar p}_{\gamma_2 \gamma_3}^0=\frac{ (\mathbf{p}_{\gamma_2,0}+\mathbf{p}_{\gamma_3,0})}{2},  
\eeq
and 
\beq
\Delta S=S_{\gamma_1} -S_{\gamma_2}+S_{\gamma_3}-S_{\gamma_4}+\delta S \, .
\eeq
Here
\beq
\delta S=(\mathbf{p}_{\gamma_1,f}-\mathbf{p}_{\gamma_2,f}
-\mathbf{p}_{\gamma_3,f}+\mathbf{p}_{\gamma_4,f})\cdot (\mathbf{r}_3-\mathbf{r}_6) /2 
\eeq
where $\mathbf{p}_{\gamma_i,f} $ stands for the final momentum of trajectory 
$\gamma_i$.

We consider here only the contribution from the diagonal terms $\gamma_1=\gamma_4$ and $\gamma_2=\gamma_3$, which leads to 
\bea\label{var4}
\delta\bar{\rho}^2(t)_{\rm diag} &=& \frac{1}{(2\pi \hbar)^{4}} {\Big \langle}
\int \prod_{i=1}^4 d{\mathbf q}_i
\sum_{\gamma_1 \left(\mathbf{q}_1\to\mathbf{q}_3,t\right)\atop \gamma_2 \left(\mathbf{q}_2\to\mathbf{q}_3,t\right) }
|D_{\gamma_1 }|^2|D_{\gamma_2 }|^2  \nonumber \\
& &\times{\rm e}^{\frac{i}{\hbar}\Delta S_{\rm d}} \rho_W\left({\mathbf q}_1,\mathbf{p}_{\gamma_1,0} \right)\rho_W\left({\mathbf q}_2,\mathbf{p}_{\gamma_{2},0} \right){\Big\rangle}_{\Delta t}.\qquad
\eea
Here $\Delta S_{\rm d}=(\mathbf{p}_{\gamma_1,f}-\mathbf{p}_{\gamma_2,f})\cdot (\mathbf{r}_3-\mathbf{r}_6)$. 
Upon applying the sum rule \cite{ref:Argaman96} this can be written as 
\beq\label{var5}
\delta\bar{\rho}^2(t)_{\rm diag}  = 
\frac{1}{(2\pi \hbar)^{2}A} {\Big \langle} \int dk \left|\left\langle{\rm e}^{-t/\tau_d}{\rm e}^{i k p^2}\right\rangle_{\mathbf{r},\mathbf{p}}\right|^2{\Big\rangle}_{\Delta t}.
\eeq
For a Gaussian initial state, the integrals can easily be performed, and for $\lambda t\gg 1$ we obtain
\beq 
\label{variance}
\delta\bar{\rho}^2(t)_{\rm diag} \approx e^{-2t/\tau_d}\frac{\sqrt{2\pi} \sigma\hbar}{A p_0} \to 0\, ,
\eeq
where $\sigma$ denotes the spatial width of the initial state 
and $p_0$ the magnitude of its mean initial momentum. Here a few remarks about are due:
(i) The result in \eq{variance} should be considered as an estimate of the leading-order $\hbar$
contribution to the variance as it is based on the diagonal approximation. The fact that it is not strictly zero in the limit $\tau_d\to \infty$ (closed system) makes us believe that there are further contributions, cancelling this term for the closed system. 
(ii) Equation (\ref{variance}) describes `mesoscopic' fluctuations of the survival probability
which turn out to be non-universal as $\delta\bar{\rho}^2(t)_{\rm diag}$ scales with the width $\sigma$
of the initial state \cite{ref:Brouwer-priv-publ}. 
(iii) Expression (\ref{variance}) may explain decay fluctuations which 
have been found from numerical calculations of the quantum decay based wave packet propagation
 \cite{ref:Waltner08}.
(iv) Furthermore, for a localized wave packet $\delta\bar{\rho}^2(t)\to 0$ as 
$\hbar\to 0$, and we have $\bar{\rho}(t)\approx \rho(t)$, recalling the result in 
Ref.\ \cite{ref:Waltner08}. 

(v) The variance (\ref{variance}) can alternatively be written as 
$\delta\bar{\rho}^2(t)_{\rm diag}\approx e^{-2t/\tau_d}/M $, where $M$ is the number of 
eigenstates of the closed system necessary to expand the initial wave function.


\section{VI. Statistics of  photofragmentation cross-sections}

Typical examples of quantum decay processes are molecular photodissociation 
\cite{ref:Baumert91,ref:Schinke93} or atomic photoionization \cite{ref:Stania05,ref:Lawley95}, where the molecule (or atom) absorbs one or
several photons such that the system is (highly) excited to an intermediate configuration coupled to
the continuum, which subsequently allows for decay, {\em i.e.} dissociation or ionization of 
the system. 

If this decay is sufficiently slow, a large portion of the complex, presumably chaotic phase space
of the excited system can be explored and the statistics of such processes are assumed to show universal 
behaviour, as described by the RMT approach developed in Refs. \cite{ref:Fyodorov98,ref:Alhassid98}. 
In these ``indirect processes''  the effective Hamiltonian of the excited molecule or atom
can be written as $H-i\Gamma/2$, where H is the Hamiltonian that represents the part of the Hamiltonian 
containing the ``binding'' potential and $\Gamma$ is a matrix describing the coupling of the system 
to N external open channels, which are the possible states of the dissociated molecule (or remaining ion). 

Semiclassical approaches to the auto-correlation function of photodissociation cross sections were still limited by the diagonal approximation used in Refs. \cite{ref:Agam00,ref:Eckhardt00}, 
which however adequately describes the leading order in $1/N$ Lorentzian profile of the 
correlation function. In Ref.\ \cite{ref:Waltner08} we briefly presented the leading off-diagonal
quantum corrections for systems with time reversal symmetry. The purpose of this section is to develop 
a semiclassical approach for quantum corrections to the  photofragmentation cross-section for 
systems with and without time reversal symmetry, including higher order corrections and finite 
Ehrenfest time effects. We follow the diagrammatic approach in Ref.\ \cite{ref:Agam00} and introduce 
2ll and 1ll contributions in order to calculate the quantum corrections. We  will see that the
form factor of the cross-section auto-correlation function can be semiclassically written as the 
the sum of the survival probability based on {\em open} trajectories in the excited system (and weighted by a factor 
which accounts for the symmetry) and the spectral form factor related to {\em periodic} orbits that 
remain trapped inside the system. 

We consider the disintegration of a molecule from its ground state $|g\rangle $ 
via photoexcitation through an intermediate excited electronic surface.

The photodissociation cross-section of the molecule, in the dipole approximation, is given by
\cite{ref:Schinke93}
\bea\label{photod1}
\sigma(E)&=& Im {\rm Tr}\{\hat A G^-(E) \}\\\nonumber &=&Im \int d{\mathbf r} \int d{\mathbf r}' A({\mathbf r},{\mathbf r'})G^-({\mathbf r}',{\mathbf r},E),
\eea
where $G^-(E)$ is the retarded Green function of the molecule, $\hat A$ is a projection operator, given by
\beq\label{photod2}
\hat A=\eta|\phi\rangle \langle \phi |, \qquad |\phi\rangle=D|g\rangle,
\eeq
where $D={\bf d}\cdot \hat{{\bf e}} $  is the projection of the electric dipole operator of 
the molecule, ${\bf d}$, on the polarization axis $\hat{{\bf e}}$ of the absorbed light,
and $\eta=E /c\hbar \epsilon_0$.

The two-point correlator of the cross-section is defined as
\beq\label{photod3}
C(\omega)\equiv
\frac{\langle \sigma(E+\hbar\omega/2)\sigma(E-\hbar \omega/2)\rangle-\langle\sigma(E)\rangle^2}{\langle \sigma(E)\rangle^2},
\eeq
where $\langle...\rangle$ denotes a local average in energy around $E$ and $\langle \sigma(E)\rangle$ 
is the mean cross-section. In the semiclassical limit $\langle \sigma(E)\rangle\approx \bar\sigma(E)$, where
\beq\label{mcf}
\bar\sigma(E)\equiv\frac{\pi}{(2\pi\hbar)^2}\int d{\mathbf r}d{\mathbf p}\, A_{W}({\mathbf r},{\mathbf
p})\delta(E-H({\mathbf r},{\mathbf p})  \, ), 
\eeq
 with
\beq\label{wfa}
A_{W}({\mathbf r},{\mathbf p})=\int d{\mathbf r'}\langle {\mathbf r}+{\mathbf r'}/2|\hat A|{\mathbf r}-{\mathbf r'}/2\rangle {\rm e}^{-i\frac{{\mathbf r'}\cdot{\mathbf p} }{\hbar}},
\eeq
the Weyl representation of the operator $\hat A$.

In the following, we consider the Fourier transform of $C(\omega)$, the cross-section form factor,
\beq\label{photod4}
Z(t)\equiv\frac{t_H}{2\pi}\int_{-\infty}^{\infty} d\omega e^{i\omega t}C(\omega).
\eeq
As $C(\omega)=C(-\omega)$ then $Z(t)$ is real and even. We consider $Z(t)$ for $t>0$ and calculate 
$C(\omega)$ from  $C(\omega)=\frac{2}{t_H}\int_{0}^{\infty} Z(t)\cos(\omega t)dt$.

In order to calculate the semiclassical expression for this quantity, we replace the exact Green function by its semiclassical counterpart \cite{ref:Gutzwiller90}, given by 
\beq\label{photod5}
G^{\rm sc}\left(\mathbf{r}',\mathbf{r},E\right)=\frac{2\pi}{(2\pi i \hbar)^{3/2}}\sum_{\tilde\gamma
\left(\mathbf{r}\to\mathbf{r}',E\right)}\tilde D_{\tilde\gamma }{\rm
e}^{\frac{i}{\hbar}\tilde S_{\tilde\gamma}(\mathbf{r},\mathbf{r'},E)} \, ,
\eeq
for a two-dimensional system, where  $\tilde D_{\tilde\gamma} =\left|\frac{\partial^2\tilde S_{\tilde \gamma}}{\partial E^2}{\rm det}\left(- \frac{\partial^2S_{\tilde \gamma}}{\partial {\mathbf r}\partial {\mathbf r}'}\right)\right|^{1/2}\exp\left(-i\frac{\pi}{2}\nu_{\tilde\gamma}\right)$, and $\nu_{\tilde\gamma}$ is the Morse index plus additional phases (see Ref.\ \cite{ref:Gutzwiller90}) and $\tilde S_{\tilde\gamma}(\mathbf{r},\mathbf{r'},E)=\int_{\mathbf{r'}}^{\mathbf{r}} {\bf p}_{\tilde\gamma}\cdot d{\bf q}_{\tilde\gamma}$ is the action integral along the trajectory $\tilde\gamma $ connecting the points $\mathbf{r'} $ and $\mathbf{r} $ with fixed energy $E$.

The semiclassical cross-section form factor \eq{photod4} is then given by
\begin{eqnarray}\label{photod6}
Z^{\rm sc}(t)&=& \frac{t_H}{8\pi\hbar^3\bar\sigma^2}{ Re}\left \langle \!\int \prod_{i=1}^4 d{\mathbf r}_i A({\mathbf r}_1,{\mathbf r}_2)A^*({\mathbf r}_3,{\mathbf r}_4)\right. \\ & & \left. \times \sum_{\tilde\gamma({\mathbf r}_1\rightarrow {\mathbf r}_2,E)\atop\tilde\gamma'({\mathbf r}_3\rightarrow {\mathbf r}_4,E)}\tilde D_{\tilde\gamma}\tilde D_{\tilde\gamma'}^* {\rm e}^{\frac{i}{\hbar}(\tilde S_{\tilde\gamma}-\tilde S_{\tilde\gamma'})}\delta\left(t-\bar t_{\gamma\gamma'}\right)\right \rangle ,\nonumber
\end{eqnarray}
where $\bar t_{\gamma \gamma'}=(t_{\tilde\gamma}+t_{\tilde\gamma'})/2$.

The term containing the action difference is a rapidly oscillating function, so due to the energy average most of the contributions will cancel out. Only trajectories with similar actions will give some contribution, which imposes conditions on the possible configuration of the points ${\mathbf r}_i$. There are two possible configurations, as depicted in \fig{fig:confi}, following the analysis in Ref.\ \cite{ref:Agam00}: (a) open trajectory (OT) contributions (that we will denote by $Z^1(t)$) where ${\mathbf r}_1 \approx {\mathbf r}_3$ and ${\mathbf r}_2 \approx {\mathbf r}_4$, or additionally, in case of time reversal symmetry,  ${\mathbf r}_1 \approx {\mathbf r}_4$ and ${\mathbf r}_2 \approx {\mathbf r}_3$ (this gives a factor of two, taking into account that in case of time reversal symmetry the eigenfunctions of $\hat H$ can be constructed to be real), (b) periodic orbit (PO) contributions ($Z^2(t)$), with ${\mathbf r}_1 \approx {\mathbf r}_2$ and ${\mathbf r}_3 \approx {\mathbf r}_4$ and both trajectories surrounding a periodic orbit.
\bfg
\centerline{ \includegraphics[width=8cm]{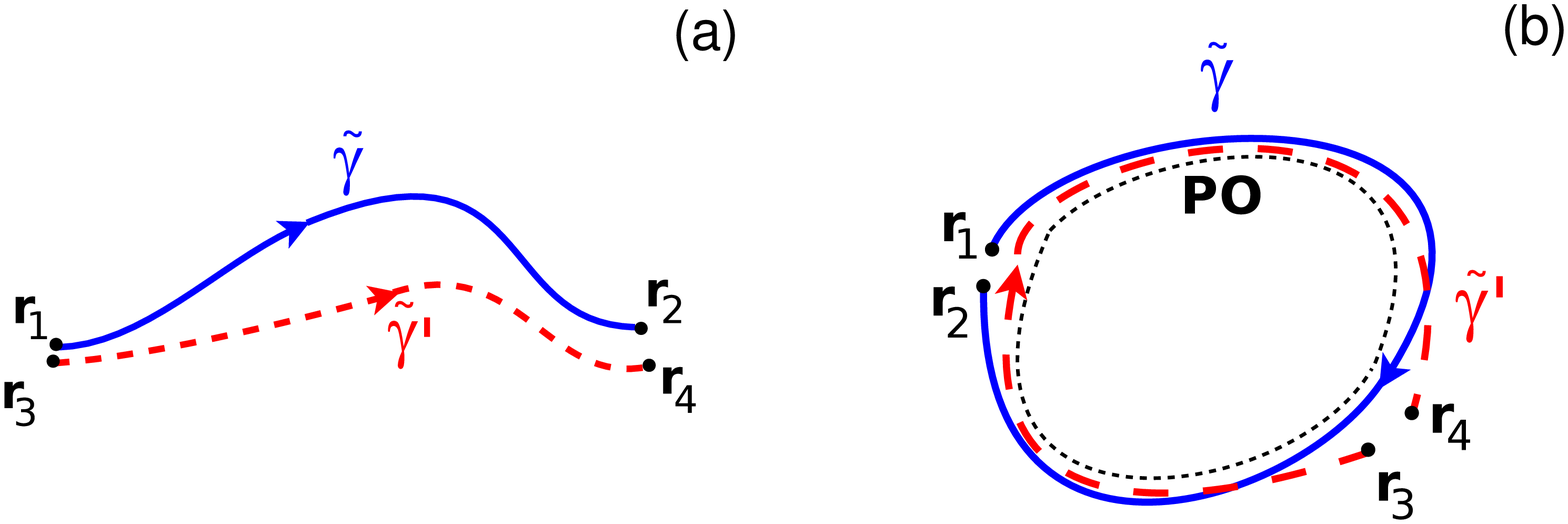}}
\caption{(color online)
Scheme of configurations giving some contribution to $Z(t)$. (a) Open trajectory  (OT) configurations counted in $Z^1(t)$. (b) Configuration with $\tilde \gamma$ and $\tilde \gamma'$ surrounding a periodic orbit (PO), contributing to $Z^2(t)$.}\label{fig:confi}
\efg

\subsection{Open trajectory contributions}

Let us consider the contributions of OT's. For this purpose we expand the contributions from 
trajectories $\tilde\gamma$ and $\tilde\gamma'$ along trajectories $\gamma$ and $\gamma'$ connecting 
${\mathbf q}=({\mathbf r}_1+ {\mathbf r}_3)/2$ and  ${\mathbf Q}=({\mathbf r}_2+ {\mathbf r}_4)/2$. Thus
\bea\label{photod7}
Z^{1}(t)&=& \frac{\kappa t_H}{8\pi\hbar^3\bar\sigma^2}{ Re}\left\langle \int d{\mathbf Q}d{\mathbf Q}'d{\mathbf q}d{\mathbf q}' \sum_{\gamma,\gamma'({\mathbf q}\rightarrow {\mathbf Q},E)}\tilde D_{\gamma}\tilde D_{\gamma'}^*\right.\nonumber\\ \nonumber & &
\times \delta(t-\bar t_{\gamma\gamma'})
 A({\mathbf q}+{\mathbf q}'/2,{\mathbf Q}+{\mathbf Q}'/2 )
 \\ & &\left.\times A^*({\mathbf q}-{\mathbf q}'/2,{\mathbf Q}-{\mathbf Q}'/2 ){\rm e}^{\frac{i}{\hbar}\Delta \tilde S_{\gamma \gamma'}}\right\rangle,
\eea
where $\Delta \tilde S_{\gamma \gamma'}=\tilde S_{\gamma}-\tilde S_{\gamma'}-({\mathbf q}'\cdot\mathbf{\bar p}_{\gamma
\gamma'}^o- {\mathbf Q}'\cdot\mathbf{\bar p}_{\gamma \gamma'}^f) $, and 
$\mathbf{\bar p}_{\gamma \gamma'}^o$  and $\mathbf{\bar p}_{\gamma \gamma'}^f$ are the averaged initial 
and final momentum of the two trajectories, respectively. Furthermore,
${\mathbf q'}={\mathbf r}_1-{\mathbf r}_2$, ${\mathbf Q'}={\mathbf r}_2-{\mathbf r}_4$ and $\kappa=1$
(or $\kappa=2$) in the absence (or presence) of time reversal symmetry.

The diagonal approximation corresponds to  $\gamma=\gamma'$. 
To evaluate these terms we invoke the sum rule from Ref.\ \cite{ref:Argaman96}, which allows us to 
write  the integrals in  Eq.\ \eq{photod7} as ${\rm e}^{-t/\tau_d(E)}\left| \int d{\mathbf q}\int
d{\mathbf p}A_W(\mathbf{q},\mathbf{p})\delta(E-H({\mathbf q},{\mathbf p}) )\right|^2 $, which in view of 
Eq.\ \eq{mcf} then gives 
\beq\label{photod9}
Z^{1, {\rm diag}}(t)=\kappa e^{-t/\tau_d} \, .
\eeq

As before, we can calculate the 2ll contribution to $Z^1(t)$ for $\kappa=2$. The double sum is replaced by the sum rule and an integral counting the encounters along $\gamma$. The classical survival probability is modified again by a factor ${\rm e}^{t_{\rm enc}/\tau_d}$. We assume that the stability amplitudes of the two trajectories are the same, so the calculation of the integral over ${\mathbf q}_i$ and ${\mathbf p}_i$ can be performed as for the diagonal approximation. Then,
\begin{eqnarray}\label{photod10}
Z^{1, \rm 2ll}(t)&=& 2\int du ds \, {\rm e}^{\frac{i}{\hbar}su} w^{\rm 2ll}(u,s,t)e^{-(t-t_{\rm enc})/\tau_d} 
\nonumber \\
&=&2e^{-t/\tau_d}\left(\frac{t^2}{2\tau_d t_H}-2\frac{t}{t_H}\right) \, .
\end{eqnarray}
As shown before in the semiclassical evaluation of double sums over OT's connecting points inside a 
system, `one-leg-loop' (1ll) diagrams have to be considered. The result for the integrals in this case is
\beq
Z^{1,\rm 1ll}\left(t\right) = 4\frac{t}{t_H}e^{-t/ \tau_d},
\eeq
cancelling the linear contribution in Eq.\ \eq{photod10}.

We note that this contribution can be written as $Z^1(t)=2\bar{\rho}(t)$, where $\bar{\rho}(t)$ is the 
mean survival probability of the state $\phi({\mathbf r})$, i.e.\
$\rho(t)=\int_A d{\mathbf r}|\phi(\mathbf r,t)|^2 $. Here the area of integration $A$ entering in the 
decay corresponds to the area confined by the binding potential.

Higher order corrections can be calculated as in Sec.~IV, and we can simply write the OT
contribution as 
\beq
\label{Z1full}
 Z^{1}=\kappa\bar{\rho}(t) \, ,
\eeq
where $\bar{\rho}(t)$ is given by Eq.\ \eq{GUE8or} for the unitary case and by Eq.\ \eq{GOE7or} for 
the orthogonal case.

\subsection{Periodic orbit contributions}

Let us now consider the contributions of diagrams such as \fig{fig:confi}b. 
We first calculate the contribution of periodic orbits to the cross section, following a similar 
procedure as for deriving the semiclassical trace formula, namely by employing the semiclassical 
Green function in the definition of $\sigma$ and expanding the actions around periodic orbits as 
shown in Ref.\ \cite{ref:Eckhardt00}
\beq\label{photod13}
\sigma^{\rm PO}(E)= \frac{1}{\hbar}Re 
\sum_{\rm j}\tilde D_{\rm j}{\rm e}^{\frac{i}{\hbar}
\tilde S_{\rm j}(E)}\; \int_0^{T_{\rm pj}}dt\, A_W({\mathbf q}_{\rm j},{\mathbf p}_{\rm j}),
\eeq
where the sum is over trapped periodic orbits ${\rm j}$, $\tilde S_j(E)=\oint_j {\bf p}\cdot d{\bf q} $ is the action integral along the periodic orbit, and $T_{\rm pj}$ refers to the 
period of the primitive periodic orbit. 
$\tilde D_{\rm j}=e^{-i\tilde\nu_{\rm j}\pi/2}/\sqrt{|{\rm Tr} M_{\rm j}-2|}$ is the stability amplitude of 
the PO together with the Maslov index $\tilde\nu_{\rm j}$, and $M_{\rm j}$ is the monodromy matrix 
describing the linearization around the PO. Almost all the long trajectories are equally distributed 
in phase space if the system is ergodic. Therefore we approximate the time integral by a corresponding
phase space average, i.e.\ $\int_0^{T_{\rm pj}}dt A_W({\mathbf q}_{\rm j},{\mathbf p}_{\rm j})\approx 
T_{\rm pj} \int d{\mathbf r}d{\mathbf p}A_{W}({\mathbf r},{\mathbf p})\delta(E-H({\mathbf r},
{\mathbf p}))/\Omega(E)$, and obtain
\begin{eqnarray}\label{photod14}
\sigma^{\rm PO}(E)&\approx& \frac{2\langle \sigma(E)\rangle }{t_H}Re 
\sum_{\rm j}T_{\rm pj} \tilde D_{\rm j}{\rm e}^{\frac{i}{\hbar}
\tilde S_{\rm j}(E)} \, .
\end{eqnarray}
We recognize here the form of the oscillatory part of the density of states. 
After substituting one finds that the contribution of periodic orbits to the cross-section form 
factor $Z^{2}(t)$ corresponds to the spectral form factor of the open system. 
Substituting Eq.\ \eq{photod14} in Eq.\ \eq{photod3} we have
\beq\label{photod15}
Z^{2}(t)= \frac{1}{t_H}{ Re}\left \langle \sum_{\rm j,j'}T_{\rm pj}T_{\rm pj'} \tilde D_{\rm j}\tilde D^*_{\rm j'}{\rm e}^{\frac{i}{\hbar}
(\tilde S_{\rm j}-\tilde S_{\rm j'})}\delta\left(t-\bar T_{\rm jj'}\right)\right \rangle,
\eeq
where $\bar T_{\rm jj'}=(T_{\rm j}+T_{\rm j'})/2$.
The expression given in Eq.\ \eq{photod15} has been calculated as an expansion in $t/t_H$ in Ref.\ \cite{ref:Kuipers07} up to 8th order for the unitary case and up to 7th order for the orthogonal case. In this context, 1ll's do not play a role, since both stretches must have a minimum time in order to surround a PO.

Summing up the semiclassical contributions to $Z^1(t)$, the decay rate \eq{Z1full}, and $Z^2(t)$, 
{\em i.e.}\ the spectral form factor $K_{\rm open}(t)$ of the open system, 
we can in general write
\beq\label{photod19}
 Z(t)=K_{\rm open}(t)+\kappa\rho(t) \, .
\eeq
Eq.\ \eq{photod19} is consistent with the result presented in Ref. \cite{ref:Fyodorov98B} for $t\ll t_H$, obtained by invoking supersymmetry techniques.
 
For the orthogonal case this is, up to 7th order in $t/t_H$,
\bea\label{ZGOE7or}
Z^{\mathrm{GOE}}(t) &=& {\rm e}^{-\frac{t}{\tau_d}}\left[2+2\frac{t}{t_H}+(N-2)\frac{t^{2}}{t_H^2}\right. \\
&& +\left(\frac{N}{3}+2\right)\frac{t^{3}}{t_H^3}+\left(\frac{5N^2}{12}-\frac{5N}{3}+\frac{8}{3}\right)\frac{t^{4}}{t_H^4} \nonumber\\
&& 
+\left(-\frac{19N^2}{60}+\frac{53N}{15}+4\right)\frac{t^{5}}{t_H^5}\nonumber \\
&& 
+\left(\frac{41N^3}{360}-\frac{N^2}{4}-\frac{101N}{15}-\frac{32}{5}\right)\frac{t^{6}}{t_H^6}\nonumber \\\nonumber
&& \left.+\left(-\frac{583N^3}{2520}+\frac{103N^2}{60}+\frac{1324N}{105}+\frac{32}{3}\right)\frac{t^{7}}{t_H^7} \right],
\eea
with $N=t_H/\tau_d$. 
For the unitary case the result reads, up to 8th order in $t/t_H$,  
\bea\label{ZGUE7or}
Z^{\mathrm{GUE}}(t) &=& {\rm e}^{-\frac{t}{\tau_d}}\left[1+\frac{t}{t_H}
 +\left(\frac{N^2}{24}-\frac{N}{6}\right)\frac{t^{4}}{t_H^4}+\frac{N^2}{24}\frac{t^{5}}{t_H^5}\right.\nonumber \\
&& 
+\left(\frac{N^2}{90}-\frac{N}{15}\right)\frac{t^{6}}{t_H^6}+\left(-\frac{N^3}{180}+\frac{N^2}{20}\right)\frac{t^{7}}{t_H^7} \nonumber \\
&& \left.+\left(\frac{N^4}{1920}-\frac{7N^3}{720}+\frac{N^2}{224}-\frac{N}{28}\right)\frac{t^{8}}{t_H^8} \right].
\eea

Returning to the auto-correlation function by taking the inverse Fourier transform, we obtain for the GOE case
\bea\label{cgoe}
C^{\rm GOE}(\Gamma)&=&4\left(\frac{1}{N}\frac{1}{1+\Gamma^2}+\frac{1}{N^2}\frac{1-\Gamma^2}{(1+\Gamma^2)^2}\right.\nonumber \\ 
&&\left.+\frac{(N-2)}{N^3}\frac{1-3\Gamma^2}{(1+\Gamma^2)^3}+ \ldots \right),
\eea
where $\Gamma=\omega\tau_d$.
The first contribution corresponds to the well known Lorentzian shaped autocorrelation function in the
regime of Ericson fluctuations, first studied by Ericson in the context of nuclear cross-sections 
in the continuum region \cite{ref:Ericson} (also experimentally observed \cite{ref:Brentano}),
and later for systems with few degrees of freedom, for which the corresponding classical scattering 
reflects irregular dynamics (`chaotic scattering') \cite{ref:Eckhardt86}. In the context of atomic
photoionization, the Lorentzian behavior has been numerically \cite{ref:Main94,ref:Madronero05} and 
experimentally \cite{ref:Stania05} studied. The first and the second term in Eq. \eq{cgoe} have been 
derived in \cite{ref:Agam00,ref:Eckhardt00}, while the third term (partly of same order $1/N^2$ 
as the second one) and higher order quantum corrections to $C(\Gamma)$ can be semiclassically assigned 
to off-diagonal loop contributions.

For the unitary case the auto-correlation function reads
\bea\label{cgue}
C^{\rm GUE}(\Gamma)&=&2\left(\frac{1}{N}\frac{1}{1+\Gamma^2}+\frac{1}{N^2}\frac{1-\Gamma^2}{(1+\Gamma^2)^2}\right.\nonumber \\ 
&&\left.+\frac{(N-4)}{N^4}\frac{(1-10\Gamma^2+5\Gamma^4)}{(1+\Gamma^2)^5}+...\right).
\eea
Eqs.\ \eq{cgoe} and \eq{cgue} are consistent with RMT results for indirect processes performed 
in Ref.\ \cite{ref:Alhassid98} and with their expansion in powers of $t/t_H$ conjectured in 
Ref.\ \cite{ref:Gorin05}. In the following section  we extend our approach beyond the RMT limit.


\section{VII. Ehrenfest time effects in photofragmentation statistics}

The Ehrenfest time $\tau_E$ \cite{ref:Chirikov} separates the short-time quantum dynamics, where quantum
wave packets follow the corresponding classical one, from a long-time regime of delocalized waves, 
where the dynamics is dominated by wave interference. Effects of this additional time scale
have been recently considered for stationary processes involving time integration, among others,
in Ref.\ 
\cite{ref:Aleiner96,ref:Yevtushenko00,ref:Adagideli03,ref:Ehrenfest2,ref:Adagideli02,ref:Brouwer06,ref:Jacquod06}. 
In Refs.\ \cite{ref:Brouwer06B,ref:Schomerus04} it was pointed out that $\tau_E$-signatures should 
be even more noticeable in the time domain. In Ref.\ \cite{ref:Waltner08} the $\tau_E$-dependence of 
the leading quantum correction to the survival probability was calculated and provided an explanation for
significant deviations of numerical quantum results in the semiclassical regime from the RMT limit. This motivates us to extend our study to $\tau_E$-effects in
the statistics of photodissociation cross-sections.

We follow the approach introduced in Ref.\ \cite{ref:Brouwer06B}, for the spectral form factor, 
to calculate the Ehrenfest time dependence of the respective  leading quantum corrections.
However we distinguish, as in Ref.\ \cite{ref:Jacquod06}, between the Ehrenfest time of the 
closed system
\beq
\label{Ehrenf-open}
\tau_E^{\rm c}\simeq\lambda^{-1}\ln({\mathcal L}/\lambda_B) \, ,
\eeq
where ${\mathcal L}$ is the typical system size and $\lambda_B$ the de Broglie wavelength,
and the open system Ehrenfest time,
\beq
\label{Ehrenf-closed}
\tau_E^{\rm o}\simeq\lambda^{-1}\ln(w^2/({\mathcal L}\lambda_B)) \, ,
\eeq
related to the width $w$ of the opening (here $w$ corresponds to the number of fragmentation channels 
times the de Broglie wavelength).

Let us consider the first (off-diagonal) quantum correction to the correlation 
function $C(\omega)$ coming from open trajectories:
\beq\label{z1te}
C^{1,\rm 2ll}_{\tau_E}(\omega)=\frac{2}{t_H} Re \int_{0}^\infty Z^{1,\rm 2ll}(t){\rm e}^{-i\omega t}dt.
\eeq
\bfg
\centerline{ \includegraphics[width=6cm]{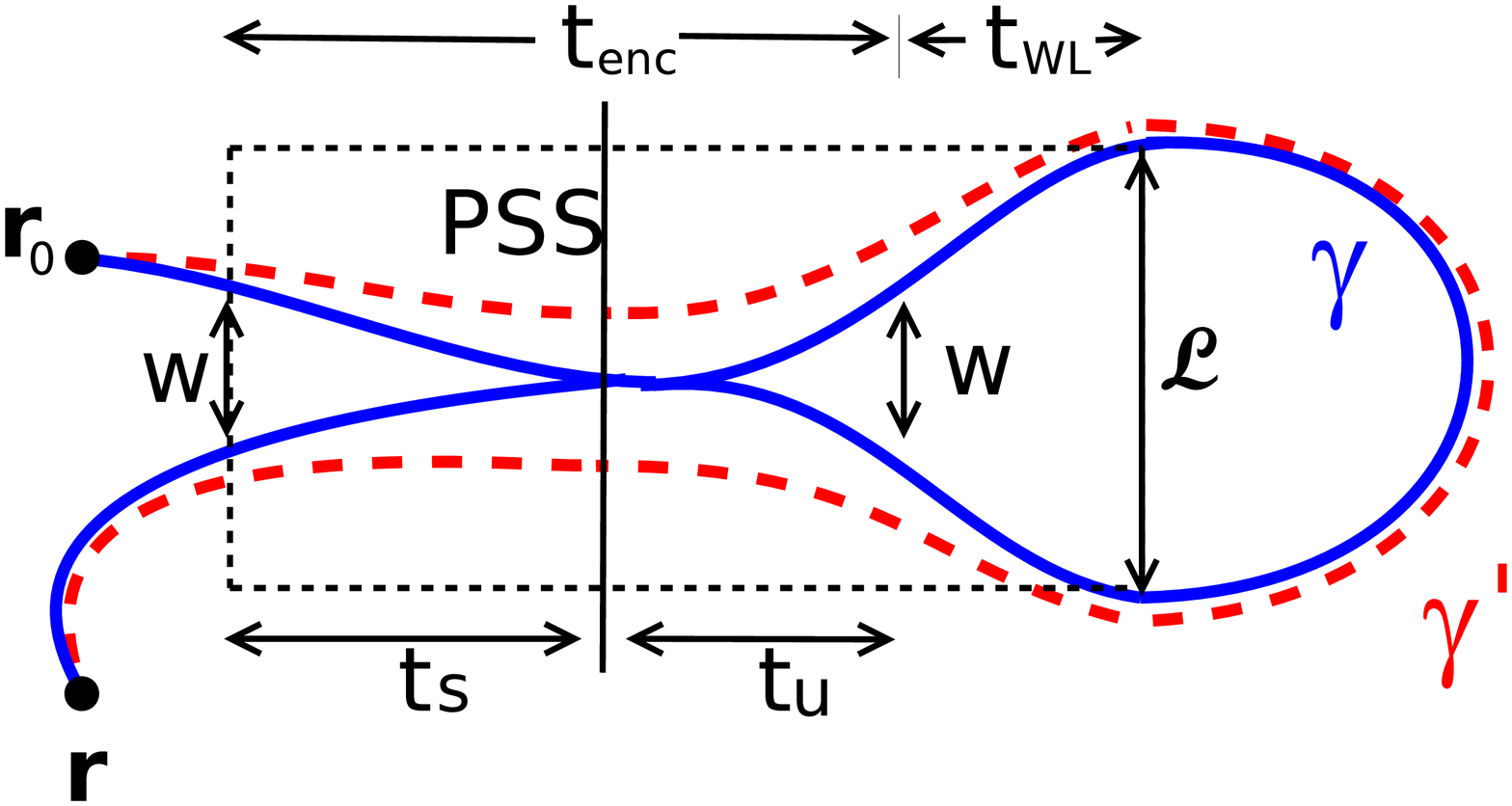}}
\caption{(color online) Sketch of the 2ll for the semiclassical approximation with finite Ehrenfest times.
}\label{fig:loopste}
\efg
As pointed out in Ref.\ \cite{ref:Waltner08} the densities should be multiplied by a Heaviside 
function ensuring that the contribution exists only for times larger than the encounter time. 
Only trajectories that are closer than a distance $w$ to themselves will have an enhanced probability 
of staying. Correlated trajectories should come closer to themselves than a distance $c^2$ in phase 
space related to the opening, i.e., we place the PSS only in the region were the stretches are closer 
than a distance $w$ in configuration space, see \fig{fig:loopste}. 
Moreover, on the right hand side of the encounter, the stretches should separate at least a distance 
${\mathcal L}$ in order to close themselves. This is because the two almost parallel momenta at the 
encounter have to grow until they are in exactly opposite directions, which requires that the stretches 
are no longer lineari\-zable along each other and therefore should be separated by a distance 
comparable to the system size. The duration of the trajectory should then be at least 
$2t_{\rm enc}+2t_{WL}$, where
\beq
\label{Ehrenf-wl}
t_{\rm WL}=\lambda^{-1}\ln(\mathcal{L}/w)
\eeq
is the time it takes for the stretches to be separated by a distance ${\mathcal L}$ when they are 
initially separated by a distance $w$. The weight function is slightly modified by this minimal 
time and by ensuring that the time is long enough in order to have such an encounter. Thus
\beq\label{w2llte}
w^{\rm 2ll}(u,s,t)=
\frac{(t-2(t_{\rm enc}+t_{\rm WL}))^2}{2\Omega t_{\rm enc}}\theta(t-2t_{\rm enc}-2t_{\rm WL}) \, ,
\eeq
and the classical survival probability is modified by ${\rm e}^{t_{\rm enc}/\tau_d}$. 
In Appendix B the evaluation of the integral can be found, together with the calculation for the 1ll case.
The total contribution can then be written as
\beq\label{tedc11s}
C^{1,\rm 2ll+1ll}_{\tau_E}(\omega)=\frac{4}{N^2}{\rm e}^{-\frac{\tau_E^c}{\tau_d}}Re
\left\{\frac{(1-i\Gamma)^3}{(1+\Gamma^2)^3}{\rm e}^{-2i\omega\tau_E^e}\right\} \, ,
\eeq
where $\Gamma=\omega\tau_d$, $N=t_H/\tau_d$ and $2\tau_E^e=\tau_E^{\rm c}+\tau_E^{\rm o}$.
Taking the Fourier transform this corresponds to a dependence in $Z^1(t)$ as
\beq\label{z1teF}
Z^{\rm 1,2ll+1ll}_{\tau_E}(t) =e^{-t/\tau_d}e^{\tau_E^o/\tau_d}\frac{(t-2\tau_E^e)^2}{\tau_dt_H}\theta(t-2\tau_E^e),
\eeq
consistent with \cite{ref:Waltner08} for the decay. Here we see two competing effects, 
on one hand if the Ehrenfest time is too large, loops can not be formed, 
$\theta(t-2\tau_E^e)=0$,  and there are no quantum contributions. On the other hand, if the time 
is long enough so that the loops can occur, i.e.\ if $t>2\tau_E^e$, the probability of staying 
is enhanced by a factor $e^{\tau_E^o/\tau_d}$ compared to generic orbits, revealing the enhanced 
classical survival probability due to the encounter. In the energy domain, the auto-correlation 
function $C(\omega)$, Eq. \eq{tedc11s}, shows an exponential suppression of quantum effects depending
on the Ehrenfest time of the closed system, similar to the exponential suppression of weak localization 
in transport in mesoscopic systems 
\cite{ref:Aleiner96,ref:Yevtushenko00,ref:Adagideli03,ref:Ehrenfest2,ref:Brouwer06,ref:Jacquod06},
while additionally oscillations in $\omega$ with a period given by $\tau_E^e$ are expected.

Let us consider now the Ehrenfest time dependence of the first quantum correction to $C^{2}(\omega)$. A calculation of the Ehrenfest time dependence of the spectral form factor of closed systems was performed in Ref.\ \cite{ref:Brouwer06B}. We follow here a similar approach, taking into account now the opening of the system, and the two different Ehrenfest time scales. In this situation the stretches are required to be separated by a distance $\mathcal{L}$ also on the left and right hand side of the encounter. Therefore the minimal time for the orbits is $2t_{\rm enc}+4t_{\rm WL}$.

\bfg
\centerline{ \includegraphics[width=6cm]{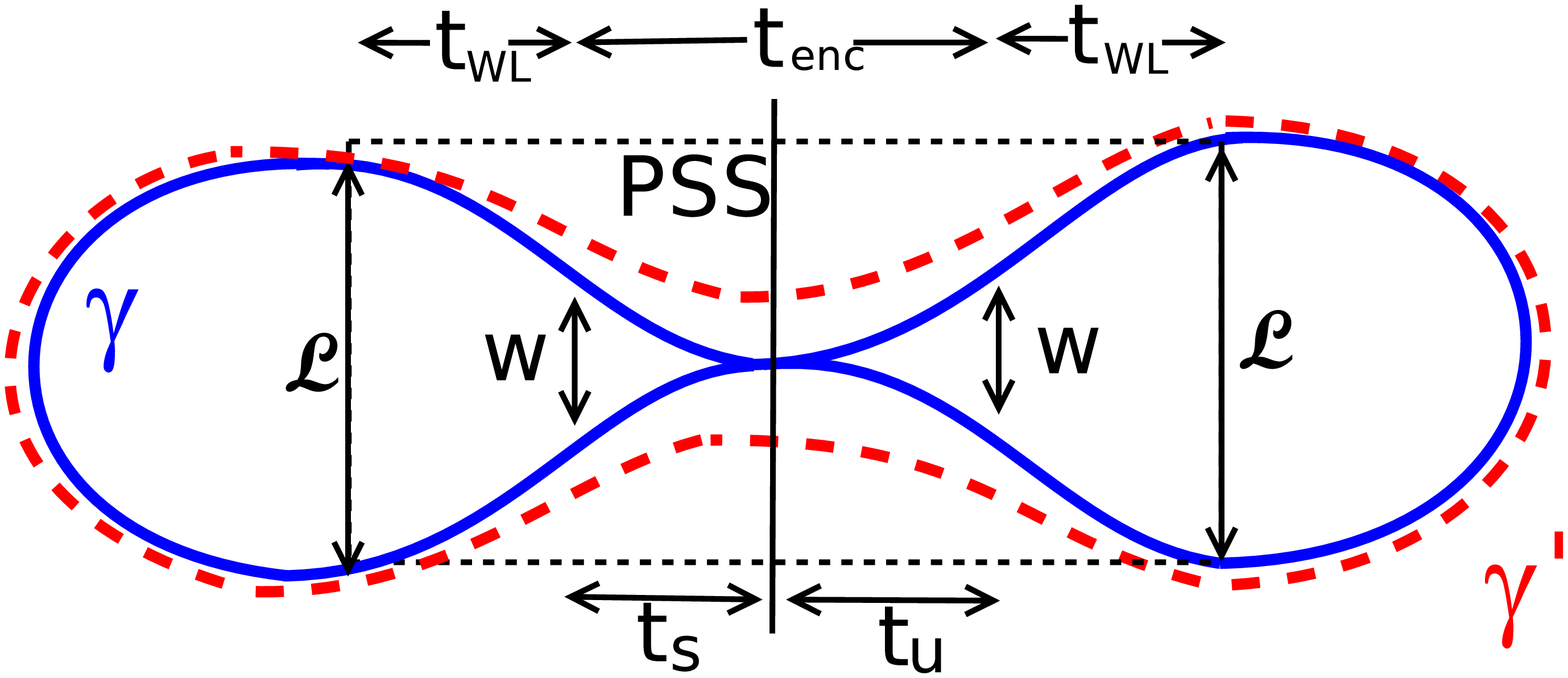}}
\caption{(color online) Sketch of a periodic orbit with a self-crossing for a finite $\tau_E$.
}\label{fig:poloopste}
\efg

The first quantum correction to the spectral form factor results from orbits sketched in 
\fig{fig:poloopste} \cite{ref:Sieber01} denoted in the following by $(2)^1$. The corrected 
weight function is then given by 
\beq
w^{(2)^1}(u,s,t)=\frac{t(t-2t_{\rm enc}-4t_{\rm WL})}{2\Omega t_{\rm enc}}\theta(t-2t_{\rm enc}-4t_{\rm
WL}) \, .
\eeq
The contribution to the autocorrelation function, after shifting the time integration by $2t_{\rm enc}$, 
can be written as 
\bea
C^{\rm 2,(2)^1}_{\tau_E}(\omega)&=&\frac{4}{t_H^3}Re\int_{4t_{\rm WL}}^\infty {\rm e}^{-(1+i\omega\tau_d)t/\tau_d} \nonumber\\ && \times (t-4t_{\rm WL})I^{\rm (2)^1}(\omega,t),
\eea 
with
\bea
I^{\rm (2)^1}(\omega,t)&=&\frac{1}{\pi\hbar}\int_0^c du\int_0^c ds {\rm e}^{\frac{i}{\hbar}us}\frac{(t+2t_{\rm enc})^2}{t_{\rm enc}}\nonumber \\ && \times{\rm e}^{-(1+2i\omega \tau_d)t_{\rm enc}/\tau_d}.
\eea
The integrals can be performed  as before, yielding
\bea\label{tedc2s}
C^{\rm 2,(2)^1}_{\tau_E}(\omega)&=&\frac{8{\rm e}^{(\tau_E^o-2\tau_E^c)/\tau_d}}{N^3}Re\left[{\rm e}^{-2i\omega \tau_E^c}\!\left( \frac{(1-2i\Gamma)}{(1+i\Gamma)^4}\right.\right.\nonumber\\ &&\left.\left.-\frac{4i\omega \tau_E^c}{(1+i\Gamma)^3}-\frac{2 \tau_E^{c \, 2}(1+2i\Gamma)}{\tau_d^2(1+i\Gamma)^2}\right) \right],\eea
where $\Gamma=\omega\tau_d$ again.
Taking the Fourier transform, the result for the spectral form factor of the open system is 
\bea \label{z2teF}
Z_{\tau_E}^{\rm 2,(2)^1}(t)&=&{\rm e}^{-t/\tau_d}{\rm e}^{\tau_E^o/\tau_d}\theta(t-2\tau_E^c)
\nonumber\\ && 
\times \left[-2\frac{t^2}{t_H^2}\left(1+\frac{\tau_E^c}{\tau_d}\right)+\frac{t^3}{\tau_d t_H^2} \right]
 \, .
\eea
If $\tau_d\to \infty$ and the system is closed, Eq.\ \eq{z2teF} is consistent with Ref.\ 
\cite{ref:Brouwer06B}. Similarly as for Eq.\ \eq{z1teF} the step function ensures that only trajectories 
longer than $2\tau_E^c$ give a contribution, which are larger than $2\tau_E^e$ since the orbits have to 
close themselves. For those orbits the contribution is enhanced by ${\rm e}^{\tau_E^o/ \tau_d}$, 
again showing the enhanced survival probability for periodic orbits with a self-encounter. 
As in Eq.\ \eq{tedc11s}, Eq.\ \eq{tedc2s} shows that the quantum corrections in the cross-section 
autocorrelation function are exponentially suppressed due to the minimal time that self-encounters require. 
In the case of periodic orbits, the suppression is stronger (since $\tau_E^c>\tau_E^o$).


\section{Conclusions and Outlook}

We have presented a detailed semiclassical analysis of the quantum survival probability and of 
photofragmentation cross-section statistics, including higher order corrections. We have
demonstrated how interference contributions associated with certain trajectory pairs provide
the key to understanding and deducing quantum corrections to the leading classical features 
in chaotic decay.
We have seen in the case of the survival probability that the initial semiclassical treatment 
introduced in Ref.\ \cite{ref:Waltner08} for localized wave packets can be extended to non-localized 
ones by assuming a local time average, which allows us to treat in the arising double sums of
trajectories pairs only those that are correlated. Apart from the standard off-diagonal contributions, 
it proves necessary to include further, so-called one-leg-loop, diagrams in order to recover unitary, 
expressed via the normalization of the wave function when the system is closed. Trajectories with 
multiple encounters of several stretches lead to higher order corrections for systems with and without 
time reversal symmetry, for which again it is necessary to take into account the corresponding
one-leg-loops as well as diagrams where both the initial and final points are inside encounter 
regions (which are not the same). Taking into account all the different allowed structures, 
depending on the general symmetries of the problem, we can reproduce RMT-type results presented in 
Ref.\ \cite{ref:Savin97}, where the survival probability was calculated using supersymmetry 
techniques. Moreover, our approach can be further extended to also 
include systems with spin-orbit interaction, which corresponds to the symplectic RMT ensemble. We have also considered mesoscopic  survival probability fluctuations
through their variance and could explicitly show that they are non-universal, that is, that
the variance depends on the spatial width of the initial (coherent) state.

In the second part of the paper, we presented in detail an application of this approach to 
a different field, namely photodissociation and photoionization processes.
We considered correlations in frequency of photofragmentation cross-sections. Its Fourier
transform, the corresponding photofragmentation form factor can be semiclassically expressed
as the sum of (twice) the survival probability, related to open trajectories, and 
the spectral form factor of the open system, related to the set of periodic orbits that are 
trapped in the open system. We have semiclassically computed the  photofragmentation form factor
to high order in $t/t_H$ and moreover considered  Ehrenfest time effects.

According to previous numerical results \cite{ref:Waltner08} there are clear indications for
the importance of Ehrenfest-time effects in decay processes, leading to a shift in time of the 
quantum corrections. In the context of photofragmentation, we have shown here that quantum 
corrections of the photodissociation form factor are also distinctly shifted in time (with a 
stronger shift for periodic orbit contributions). This time shift translates into 
an exponential suppression of quantum effects in the cross-section correlator, if the Ehrenfest 
time is comparable to the typical life time of the intermediate atomic or molecular resonant
states in the fragmentation process. Our semiclassical results also predict
a frequency modulation of the correlator with period given by the Ehrenfest time.

The semiclassical approach developed here to treat decay processes can be extended
 to address other quantities where semiclassics so far was limited
by the diagonal approximation. One example is the problem of the Loschmidt echo 
or fidelity, respectively, where a semiclassical treatment along similar lines as the one 
presented here allows one to calculate quantum corrections to the fidelity decay \cite{ref:gutkin-etal08}.

The present approach is still limited to times smaller than the Heisenberg time. An extension
to longer times beyond $t_H$ remains as a challenging open problem of semiclassics for
open quantum systems.


\section{Acknowledgments}

\noindent We thank P.\ Brouwer, A.\ Goussev, C.~Petitjean and D.~Savin for helpful discussions. 
We acknowledge funding by DFG under GRK 638. 
\appendix


\section{Appendix A: Higher order contributions to the decay rate in the orthogonal case}

\begin{table}[thb]
\centering
\begin{tabular}{|c|c|c|c|c|c|c|}
\hline
$\v$&$L$&$V$&$N(\v)$&$\mathcal{N}_{l_1,l_2}(\v)$&$\mathcal{N}_{l_1,l_V}(\v)$&$\mathcal{N}_{l_{V-1},l_V}(\v)$\\
\hline
$(2)^{1}$&2&1&1&&&\\
\hline
$(2)^{2}$&4&2&5&4&&\\
$(3)^{1}$&3&1&4&&&\\
\hline
$(2)^{3}$&6&3&41&36&&\\
$(2)^{1}(3)^{1}$&5&2&60&40&&\\
$(4)^{1}$&4&1&20&&&\\
\hline
$(2)^{4}$&8&4&509&468&&\\
$(2)^{2}(3)^{1}$&7&3&1092&228&672&\\
$(2)^{1}(4)^{1}$&6&2&504&296&&\\
$(3)^{2}$&6&2&228&148&&\\
$(5)^{1}$&5&1&148&&&\\
\hline
$(2)^{5}$&10&5&8229&7720&&\\
$(2)^{3}(3)^{1}$&9&4&23160&8220&12256&\\
$(2)^{2}(4)^{1}$&8&3&12256&1884&7480&\\
$(2)^{1}(3)^{2}$&8&3&10960&5024&&3740\\
$(2)^{1}(5)^{1}$&7&2&5236&2696&&\\
$(3)^{1}(4)^{1}$&7&2&4396&2696&&\\
$(6)^{1}$&6&1&1348&&&\\
\hline
$(2)^{6}$&12&6&166377&158148&&\\
$(2)^{4}(3)^{1}$&11&5&579876&266040&265056&\\
$(2)^{3}(4)^{1}$&10&4&331320&93456&186160&\\
$(2)^{2}(3)^{2}$&10&4&443400&41792&249216&93080\\
$(2)^{2}(5)^{1}$&9&3&167544&19872&98712&\\
$(2)^{1}(3)^{1}(4)^{1}$&9&3&280368&49576&66240&98712\\
$(3)^{3}$&9&3&41792&33120&&\\
$(2)^{1}(6)^{1}$&8&2&65808&30208&&\\
$(3)^{1}(5)^{1}$&8&2&52992&30208&&\\
$(4)^{2}$&8&2&24788&15104&&\\
$(7)^{1}$&7&1&15104&&&\\
\hline
\end{tabular}
\caption{The number of trajectory pairs and the number linking certain encounters for systems with time reversal symmetry.}
\label{goenumbers}
\end{table}

Table \ref{goenumbers} allows us to obtain the following semiclassical corrections to
$\rho^{\rm cl} = \rme^{-\frac{t}{\tau_d}}$  for the orthogonal case:

\beq
\bar\rho_{1}(t)= \frac{\rme^{-\frac{t}{\tau_d}}}{t_H}\left(\frac{t^{2}}{2\tau_d}\right),
\eeq

\beq
\bar\rho_{2}(t)= \frac{\rme^{-\frac{t}{\tau_d}}}{t_H^2}\left(-\frac{t^{3}}{3\tau_d}+\frac{5t^{4}}{24\tau_d^{2}}\right),
\eeq

\beq
\bar\rho_{3}(t) = \frac{\rme^{-\frac{t}{\tau_d}}}{t_H^3}\left(\frac{t^{4}}{3\tau_d}-\frac{11t^{5}}{30\tau_d^2}+\frac{41t^{6}}{720\tau_d^3}\right),
\eeq

\beq
\bar\rho_{4}(t) = \frac{\rme^{-\frac{t}{\tau_d}}}{t_H^4}\left(-\frac{2t^{5}}{5\tau_d}+\frac{7t^{6}}{12\tau_d^{2}}-\frac{29t^{7}}{168\tau_d^{3}}+\frac{509t^{8}}{40320\tau_d^{4}}\right),
\eeq
\begin{eqnarray}
\bar\rho_{5}(t) &=& \frac{\rme^{-\frac{t}{\tau_d}}}{t_H^5}\left(\frac{8t^{6}}{15\tau_d}-\frac{14t^{7}}{15\tau_d^{2}}+\frac{31t^{8}}{80\tau_d^{3}}\right. \nonumber \\
&& \qquad\qquad\left. -\frac{271t^{9}}{5040\tau_d^{4}}+\frac{2743t^{10}}{1209600\tau_d^{5}}\right),\end{eqnarray}
\begin{eqnarray}
\bar \rho_{6}(t) &=& \frac{\rme^{-\frac{t}{\tau_d}}}{t_H^6}\left(-\frac{16t^{7}}{21\tau_d}+\frac{5099t^{8}}{3360\tau_d^{2}}-\frac{4469t^{9}}{5670\tau_d^{3}}+\frac{437t^{10}}{2800\tau_d^{4}}\right. \nonumber \\
&& \qquad\qquad\left. -\frac{28001t^{11}}{2217600\tau_d^{5}}+\frac{55459t^{12}}{159667200\tau_d^{6}}\right).
\end{eqnarray}


\section{Appendix B: Ehrenfest time dependence of the leading quantum correction to 
 the cross-section correlation $C(\omega)$}

Substituting the expressions \eq{Ehrenf-wl} and \eq{w2llte} into Eq.~\eq{z1te} and shifting the time integral 
by $2t_{\rm enc}+2t_{\rm WL}$ we have for the 2ll correction to the photo cross-section correlation
\beq\label{z1teB}
C^{1,\rm 2ll}_{\tau_E}(\omega)=\frac{4}{t_H^2} Re \int_{0}^\infty \,t^2\, {\rm e}^{-(1+i\omega\tau_d) (t+2t_{\rm WL})/\tau_d} I^{\rm 2ll}(\omega)  dt
\eeq
with
\beq
I^{\rm 2ll}(\omega)=\frac{1}{\pi\hbar}\int_0^{c}du\int_0^{c}ds \frac{{e}^{\frac{i}{\hbar}us}}{ t_{\rm enc}}{\rm e}^{t_{\rm enc}/\tau_d}{\rm e}^{-2(1+i\omega\tau_d)t_{\rm enc}/\tau_d}.
\eeq
Upon change of variables, $x=us/c^2$ and $\sigma=c/u$, we obtain
\beq
I^{\rm 2ll}(\omega)=\frac{r\lambda}{\pi}\int_0^{1}dx
\cos(rx)x^{(1+2i\omega\tau_d)/(\lambda\tau_d)} \, ,
\eeq
where $r=c^2/\hbar$, and the integral over $\sigma$ has already been performed. We compute the
remaining integral by partial integration, neglecting highly oscillating terms in the limit 
$\hbar\to 0$ while keeping $\tau_E^{o}/\tau_d$ (Eq.~\eq{Ehrenf-open}) and $\tau_E^{c}/\tau_d$ 
(Eq.~\eq{Ehrenf-closed}) finite. We find
\beq
I^{\rm 2ll}(\omega)=-\frac{(1+2i\omega\tau_d)}{2\tau_d}{\rm e}^{\tau_{E}^o/\tau_d}e^{-2(1+i\omega\tau_d)\tau_{E}^o/\tau_d},
\eeq
with Ehrenfest time $\tau_E^{\rm o}\!=\!\lambda^{-1}\!\ln(c^2/\hbar)$. 
Then Eq.~\eq{z1teB} gives 
\beq
\label{C12ll}
C^{1,\rm 2ll}_{\tau_E}(\omega)=-\left(\frac{2\tau_d}{t_H}\right)^2
{\rm e}^{-\frac{\tau_E^c}{\tau_d}}Re\left[\frac{1+2i\omega\tau_d}{(1+i\omega\tau_d)^3}{\rm e}^{-2i\omega\tau_E^e}\right].
\eeq
Here we used that from the definitions of $\tau_E^{\rm o}$ and $t_{\rm WL}$ follows  
$\tau_E^{\rm o}+2t_{\rm WL}=\tau_E^{\rm c}$, and we introduced
$\tau_E^e=(\tau_E^{\rm c}+\tau_E^{\rm o})/2$.

A corresponding calculation can be performed for the 1ll case, where 
\beq\label{z1teBp}
C^{1,\rm 1ll}_{\tau_E}(\omega)=\frac{16}{t_H^2} Re \int_{0}^\infty \,t\,{\rm e}^{-(1+i\omega\tau_d)(t+2t_{WL})/\tau_d} I^{\rm 1ll}(\omega) dt
\eeq
with
\bea
I^{\rm 1ll}(\omega)&=&\frac{1}{\pi\hbar}\int_0^{c}du\int_0^{c}ds\int_0^{\lambda^{-1}\ln(c/|s|) } dt' \frac{{e}^{\frac{i}{\hbar}us}}{ t_{\rm enc}}
\\ \nonumber && \times {\rm e}^{t_{\rm enc}/\tau_d}{\rm e}^{-2(1+i\omega\tau_d)t_{\rm enc}/\tau_d},
\eea
and $t_{\rm enc}=t'+\lambda^{-1}\ln(c/|u|) $. 
With the change of variables  $x=us/c^2$ and $\sigma=c/u$ and $t''=t'+\lambda^{-1}\ln(c/|u|)$, we obtain
\beq
I^{\rm 1ll}(\omega)=-\frac{\lambda r \tau_d}{\pi(1+2i\omega\tau_d)}\int_0^{1}dx
\cos(rx)x^{\frac{1}{\lambda\tau_d}(1+2i\omega\tau_d)}\ .
\eeq
The integral can be evaluated as before, neglecting  higher-order terms. 
Together with $C^{1,\rm 2ll}_{\tau_E}(\omega)$ from Eq.~\eq{C12ll} we find
\beq
C^{1,\rm 2ll+1ll}_{\tau_E}(\omega)=4\frac{\tau_d^2}{t_H^2}{\rm
e}^{-\frac{\tau_E^c}{\tau_d}}Re\left[\frac{(1-i\omega\tau_d)^3}{(1+(\omega\tau_d)^2)^3}{\rm
e}^{-2i\omega\tau_E^e}\right] \, ,
\eeq
which corresponds to Eq.~\eq{tedc11s}.



\begin{thebibliography}{00}

\bibitem{ref:Bacher99}
G.~Bacher, R.~Weigand, J.~Seufert, V. D.~Kulakovskii, N. A.~Gippius, A.~Forchel, K.~Leonardi, and D.~Hommel, Phys.\ Rev.\ Lett.\ {\bf 83}, 4417 (1999).


\bibitem{ref:Kumar98}
R.~Kumar, A. S.~Vengurlekar, A.~Venu Gopal, T.~M\'elin, F.~Laruelle, B.~Etienne, and J.~Shah, Phys.\ Rev.\ Lett.\ {\bf 81}, 2578 (1998).  

\bibitem{ref:Stania05}
G.~Stania and H.~Walther, Phys.\ Rev.\ Lett. {\bf 95}, 194101 (2005).


\bibitem{ref:Baumert91}
T.~Baumert, M.~Grosser, R.~Thalweiser, and G.~Gerber, Phys. \ Rev. \ Lett.\ {\bf 67}, 3753 (1991).


\bibitem{ref:Schinke93}
R.~Schinke, {\it Photodissociation Dynamics} (Cambridge University Press, Cambridge, 1993).


\bibitem{ref:atombilliard}
V. Milner, J. L. Hanssen, W. C. Campbell, and M. G. Raizen, Phys. \ Rev. \ Lett.\ {\bf 86}, 1514 (2001);
N.~Friedman, A.~Kaplan, D.~Carasso, and N.~Davidson,  {\em ibid}, 1518 (2001);

\bibitem{ref:microcav}
J.~U.~N\"ockel and A.~D.~Stone, Nature (London) {\bf 385}, 45 (1997);
T.~Harayama, P.~Davis, and K. S.~Ikeda, Phys.\ Rev.\ Lett.\ {\bf 90}, 063901 (2003);
W.~Fang, A.~Yamilov, and H.~Cao, Phys.\ Rev.\ A {\bf 72}, 023815 (2005);
J.~Wiersig and M.~Hentschel, Phys.\ Rev.\ Lett.\ {\bf  100}, 033901 (2008).

\bibitem{ref:Casati97}
G.~Casati, G.~Maspero, and D. L.~Shepelyansky, Phys.\ Rev.\ E {\bf 56}, R6233 (1997); 
G.~Casati, I.~Guarneri, and G.~Maspero, Phys.\ Rev.\ Lett. {\bf 84}, 63 (2000); G.~Casati, G.~Maspero, and D. L.~Shepelyansky, Phys.\ Rev.\ Lett. {\bf 82}, 524 (1999).

\bibitem{ref:Frahm97}
K. M.~Frahm, Phys.\ Rev.\ E {\bf 56}, R6237 (1997).


\bibitem{ref:Savin97}
D. V.~Savin and V. V.~Sokolov, Phys.\ Rev.\ E {\bf 56}, R4911 (1997); 
D. V.~Savin and H.-J.~Sommers, Phys. \ Rev. \ E {\bf 68}, 036211 (2003).


\bibitem{ref:Puhlmann05}
M.~Puhlmann, H.~Schanz, T.~Kottos, and T.~Geisel, Europhys.\ Lett. {\bf 69}, 313 (2005).

\bibitem{ref:Waltner08}
D.~Waltner, M.~Guti\'errez, A. Goussev, and K. Richter, Phys. Rev. Lett. {\bf 101}, 174101 (2008).

\bibitem{ref:Sieber01}
M.~Sieber and K.~Richter, Phys.\ Scr.\ {\bf T90}, 128 (2001); 
M.~Sieber, J.~Phys. A {\bf 35}, L613 (2002).


\bibitem{ref:chaos-general}
S.~M\"uller, Eur.~Phys.~J.~B {\bf 34}, 305 (2003); 
M.~Turek and K.~Richter, J.~Phys.~A {\bf 36}, L455 (2003);
D.~Spehner, J.~Phys.~A {\bf 36}, 7269 (2003). 

%
\bibitem{ref:Heusler04}
S.~M\"uller, S.~Heusler, P.~Braun, F.~Haake, and A.~Altland, Phys.\ Rev.\ Lett.\ {\bf 93}, 014103 (2004); 
S.~M\"uller, S.~Heusler, P.~Braun, F.~Haake, and A.~Altland, Phys.\ Rev.\ E\ {\bf 72}, 046207 (2005);
S.~Heusler, S.~M\"uller, A.\ Altland, P.~Braun, and F.~Haake, Phys.~Rev.~Lett.~{\bf 98} 044103 (2007);
S.~M\"uller, {\it Periodic orbit approach to universality in quantum chaos}, PhD thesis, arXiV:nlin/0512058v1 (2005).


\bibitem{ref:Brouwer06B}
P. W.~Brouwer, S.~Rahav, and C.~Tian, Phys.\ Rev.\ E {\bf 74}, 066208 (2006).

\bibitem{ref:Richter02}
K. Richter and M. Sieber, Phys.\ Rev.\ Lett.\ {\bf 89}, 206801 (2002).

 \bibitem{ref:Adagideli03}
 \.{I}.~Adagideli, Phys.\ Rev.\ B {\bf 68}, 233308 (2003).


\bibitem{ref:Heusler06}
 S.~Heusler, S.~M\"uller, P.~Braun, and F.~Haake, Phys.\ Rev.\ Lett.\ {\bf 96}, 066804 (2006);
S.~M\"uller, S.~ Heusler, P.~ Braun, and F.~ Haake, New J.\ Phys.\ {\bf 9}, 12 (2007).


 \bibitem{ref:Brouwer06}
 P. W.~Brouwer and S.~Rahav, Phys.\ Rev.\ B {\bf 74}, 075322 (2006).


 \bibitem{ref:Ehrenfest2}
 S.~Rahav and P.~W.~Brouwer, Phys.\ Rev.\ Lett.~{\bf 96}, 196804 (2006).


 \bibitem{ref:Jacquod06}
 Ph.~Jacquod and R. S.~Whitney, Phys.\ Rev.\ B {\bf 73}, 195115 (2006).

\bibitem{ref:Kuipers08}
 J.~Kuipers and M.~Sieber, Phys.\ Rev.\ E {\bf 77}, 046219 (2008).

\bibitem{ref:Whitney07}
R.~S.~Whitney, Phys.\ Rev.\ B {\bf 75}, 235404 (2007).

 
 \bibitem{ref:Agam00}
 O.~Agam, Phys.\ Rev.\ E {\bf 61}, 1285 (2000).

 
 \bibitem{ref:Eckhardt00}
 B.~Eckhardt, S.~Fishman, and I.~Varga,  Phys.\ Rev.\ E {\bf 62}, 7867 (2000).

 
 \bibitem{ref:Gutzwiller90}
 M.~Gutzwiller, {\it Chaos in Classical and Quantum Mechanics} (Springer, New York, 1990).

 
 \bibitem{ref:Cucchietti04}
 F. M.~Cucchietti, H. M.~Pastawski, and R. A.~Jalabert, Phys.\ Rev. \ B {\bf 70}, 035311 (2004).

 
 \bibitem{ref:Sieber99}
 M.~Sieber, J.\ Phys.\ A {\bf 32}, 7679 (1999).

 
 \bibitem{ref:Kuipers07}
 J.~Kuipers and M.~Sieber, Nonlinearity {\bf 20}, 909 (2007).

 
 \bibitem{ref:Zaitsev05}
 O.~Zaitsev, D.~Frustaglia, and K.~Richter, Phys. Rev. Lett. {\bf 94}, 026809 (2005); Phys. Rev. B {\bf 72}, 155325 (2005);
 O.~Zaitsev and K.~Richter, Mat.\ Science-Poland {\bf 22}, 469 (2004).


 \bibitem{ref:Bolte07}
 J.~Bolte and  D.~Waltner, Phys. Rev. B {\bf 76}, 075330 (2007).


 \bibitem{ref:Bolte99}
 J.~Bolte and  S.~Keppeler, Ann. Phys. (N. Y.) {\bf 274}, 125 (1999).

\bibitem{ref:Argaman96}
 N.~Argaman, Phys.\ Rev.\ Lett. {\bf 75}, 2750 (1995); 
Phys.\ Rev.\ B {\bf 53}, 7035 (1996).



\bibitem{ref:Brouwer-priv-publ} 
    A result corresponding to \eq{variance} has been independently
    obtained by P.~W.~Brouwer (private communication).

\bibitem{ref:Lawley95}
K.~P.~Lawley (Ed.), {\it Photodissociation and Photoionization} (Wiley, New York, 1995). 

 \bibitem{ref:Fyodorov98}
 Y. V.~Fyodorov and Y.~Alhassid, Phys.\ Rev.\ A {\bf 58} R3375 (1998).

 
 \bibitem{ref:Alhassid98}
 Y.~Alhassid and Y. V.~Fyodorov, J.\ Phys.\ Chem.\ A {\bf 102}, 9577 (1998).

 
 \bibitem{ref:Fyodorov98B}
Taking in Eq. (8) of Y.~Alhassid, Y.~V.~Fyodorov, T.~Gorin, W.~Ihra, and B.~Mehlig, Phys.\ Rev.\ A {\bf 73} 042711 (2006) the limit of no direct coupling to the continuum and $t\gg t_H$ leads to Eq. \eq{photod19}. 
 
 \bibitem{ref:Ericson}
 T.~Ericson, Phys.\ Rev.\ Lett. {\bf 5}, 430 (1960); 
 Phys.\ Lett. {\bf 4}, 258 (1963); 
 Ann.\ Phys.\ {\bf 23}, 390 (1963).

 
 \bibitem{ref:Brentano}
 P.~von Brentano, J.~ Ernst, O.~H\"ausser, T.~Mayer-Kuckuk, A.~ Richter, and W.~von Witsch, Phys.\ Lett. {\bf 9}, 48 (1964).

 
 \bibitem{ref:Eckhardt86}
 B.~Eckhardt and C.~ Jung, J.\ Phys.\ A {\bf 19}, L829 (1986);
 B.~Eckhardt, J.\ Phys.\ A {\bf 20}, 5971 (1987);
 C.~Jung and H. Scholz, J.\ Phys.\ A {\bf 20}, 3607 (1987).

 
 \bibitem{ref:Main94}
 J.~Main and G.~Wunner, J.\ Phys.\ B {\bf 27}, 2835 (1994);
 V.~V.~ Flambaum, A.~A. Gribakina, and G.~F.~Gribakin, Phys. \ Rev. \ A {\bf 54}, 2066 (1996).


 \bibitem{ref:Madronero05}
 J.~Madro\~nero and A.~Buchleitner, Phys.\ Rev.\ Lett. {\bf 95}, 263601 (2005).


 \bibitem{ref:Gorin05}
 T.~Gorin,  J.\ Phys.\ A {\bf 38}, 10805 (2005).

 
 \bibitem{ref:Chirikov}
 B.~V.~Chirikov, F.~M.~Izrailev, and D.~L.~Shepelyansky, Sov.~Sci.~Rev. Sect.~C {\bf 2}, 209 (1981).

 
 \bibitem{ref:Aleiner96}
 I.~L.~Aleiner and A.~I.~Larkin, Phys.~Rev.~B {\bf 54}, 14423 (1996).

 
 \bibitem{ref:Yevtushenko00}
 O.~Yevtushenko, G.~L\"utjering, D.~Weiss, and K.~Richter, Phys.~Rev.~Lett.\ {\bf 84}, 542 (2000).

 \bibitem{ref:Adagideli02}
P.~Jacquod, \.{I}.~Adagideli, and C.W.J.~Beenakker, Phys.~Rev.~Lett.\ {\bf 89}, 154103 (2002).


 \bibitem{ref:Schomerus04}
 H.~Schomerus and J.~Tworzyd\l o,  Phys.\ Rev.\ Lett.\ {\bf 93}, 154102 (2004).


\bibitem{ref:gutkin-etal08}
B.~Gutkin, D.~Waltner, M.~Guti\'errez, J.~Kuipers, and K.~Richter, in preparation (2008).

 \end{thebibliography}
 \end{document}